\documentclass[12pt]{article}
\pdfoutput=1

\usepackage[utf8]{inputenc}
\usepackage{geometry}
\usepackage[pdfencoding=auto]{hyperref}
\usepackage{cite}
\hypersetup{pageanchor=true,citecolor=blue,urlcolor=blue,colorlinks=true,linkcolor=blue}
\usepackage[toc,page]{appendix}
\usepackage[]{graphicx}
\usepackage{amsmath}
\usepackage{mathrsfs}
\usepackage{amssymb}
\usepackage{array}
\usepackage{chngcntr}

\usepackage[shortlabels]{enumitem}

\let\OLDthebibliography\thebibliography
\renewcommand\thebibliography[1]{
    \OLDthebibliography{#1}
    \setlength{\parskip}{5pt}
    \setlength{\itemsep}{0pt plus 0.3ex}
}

\counterwithin*{equation}{section}

\usepackage{cleveref}

\newcommand{\be}{\begin{equation}}
\newcommand{\ee}{\end{equation}}
\newcommand{\bl}{\begin{align}}
\newcommand{\el}{\end{align}}
\newcommand{\bseq}{\begin{subequations}}
\newcommand{\eseq}{\end{subequations}}

\renewcommand{\l}{\lambda}
\renewcommand{\O}{\Omega}

\newcommand{\vf}{\varphi}

\newcommand{\diff}{\mathrm{d}}
\newcommand{\e}{{\rm e}}

\renewcommand{\th}{\mathop{\rm th}\nolimits}

\renewcommand{\Im}{\mathop{\rm Im}\nolimits}
\newcommand{\ch}{\mathop{\rm ch}\nolimits}
\newcommand{\sh}{\mathop{\rm sh}\nolimits}

\newcommand{\UU}{\mathcal{U}}
\newcommand{\VV}{\mathcal{V}}

\usepackage[normalem]{ulem}

\begin{document}
\begin{titlepage}

\title{\vspace{-3.0cm}
\begin{flushright}
{\small FTPI-MINN-22-24}
\end{flushright}
\vspace{-0.8cm}
\begin{flushright}
{\small UMN-TH-4133/22}
\end{flushright}
\vspace{-1.3cm}
\begin{flushright}
{\small CP3-22-43}
\end{flushright}
\vspace{1.5cm}
{\LARGE\bf
Non-perturbative production of fermionic \\ dark matter from fast preheating
}
}

\author{Juraj Klaric$^{a}$\footnote{juraj.klaric@uclouvain.be}~,
Andrey Shkerin$^{b}$\footnote{ashkerin@umn.edu}~,
Georgios Vacalis$^{c}$\footnote{georgios.vacalis@physics.ox.ac.uk}\\[2mm]
{\small\it $^a$Centre for Cosmology, Particle Physics and Phenomenology, } \\
{\small \it Université catholique de Louvain, Louvain-la-Neuve B-1348, Belgium} \\
{\small\it $^b$William I. Fine Theoretical Physics Institute, School
  of Physics and Astronomy,} \\
{\small\it University of Minnesota, Minneapolis, MN 55455, USA }\\
{\small\it $^c$Department of Physics, University of Oxford, Parks Road, Oxford OX1 3PU, UK}
}

\date{}
\maketitle

\begin{abstract}

We investigate non-perturbative production of fermionic dark matter in the early universe.
We study analytically the gravitational production mechanism accompanied by the coupling of fermions to the background inflaton field.
The latter leads to the variation of effective fermion mass during preheating and makes the resulting spectrum and abundance sensitive to its parameters.
Assuming fast preheating that completes in less than the inflationary Hubble time and no oscillations of the inflaton field after inflation, we find an abundant production of particles with energies ranging from the inflationary Hubble rate to the inverse duration of preheating.
The produced fermions can account for all observed dark matter in a broad range of parameters.
As an application of our analysis, we study non-perturbative production of heavy Majorana neutrino in the model of Palatini Higgs inflation.

\end{abstract}

\thispagestyle{empty}
\end{titlepage}

\newpage
\tableofcontents

\section{Introduction}
\label{sec:intro}

The nature of dark matter (DM) remains one of the biggest open questions of cosmology.
Most commonly it is identified with an elementary particle beyond the Standard Model.
Examples are numerous; they span over many orders of magnitude of the mass of the DM particle and explore different forms of coupling between the DM and the Standard Model and, possibly, other fields such as the inflaton field (see, e.g., \cite{Feng:2010gw,Bertone:2016nfn,Arun:2017uaw} for reviews).

An important part in proposing a DM candidate is to provide a mechanism of its production in the early universe.
Here, gravity can play an important role.
Indeed, while we cannot be certain about interactions of DM with other fields, we are certain that the DM gravitates.
Gravity provides inevitable channels of DM production that set the lower bound on its abundance.
Depending on the DM mass, the gravitational mechanism can be efficient enough to account for all observed DM.
This is true, in particular, for the non-perturbative particle production due to varying geometry of the expanding universe.
Including interactions between the DM and other fields opens other production channels and generally enhance the resulting abundance.

The non-perturbative gravitational production of massive fermions in the early universe has been extensively studied in the literature~\cite{Parker:1971pt,1975SvAL....1..179M,1978SvAL....4..111M,Audretsch:1978qu,Kuzmin:1998uv,Kuzmin:1998kk,Kuzmin:1999zk,Chung:2001cb,Chung:2011ck,Adshead:2015kza,Ema:2019yrd,Herring:2020cah}.\:\footnote{For studies of the perturbative production of fermions from the $s$-channel graviton exchange during preheating see~\cite{Haque:2021mab,Mambrini:2021zpp,Clery:2021bwz,Haque:2022kez}.}
In the conformally-flat FLRW (Friedmann--Lemaître--Robertson--Walker) metric, the production caused by the varying geometry can be described using the time-varying ``effective'' fermion mass
\begin{equation}\label{Meff1}
    M_{\rm eff}(\eta)=M a(\eta) \;,
\end{equation}
where $M$ is the ``bare'' mass and $a(\eta)$ is the scale factor depending on the conformal time $\eta$.
During the radiation-dominated stage the production is most efficient at $M\sim H$ where $H=H(\eta)$ is the Hubble rate~\cite{Parker:1971pt,1975SvAL....1..179M,1978SvAL....4..111M,Audretsch:1978qu}.
In the inflationary cosmology, the Big-Bang singularity is replaced by the quasi-de Sitter stage that is connected to the radiation-dominated stage by the period of preheating.\:\footnote{Here we do not consider alternative scenarios such as the kination stage preceding the radiation-dominated epoch~\cite{Spokoiny:1993kt,Joyce:1997fc,Pallis:2005hm,Pallis:2005bb}.}
The question then arises as to what degree the spectrum and abundance of produced particles are sensitive to the duration and details of preheating.

Possible non-gravitational couplings of the dark fermion can also contribute to its effective mass and, hence, to its production in the post-inflationary universe.
For example, the Yukawa interaction between the Dirac or Majorana fermion $\Psi$ and the inflaton $\Phi$ fields,
\begin{equation}\label{Int1}
    \mathcal{L}_{int.}=c\:\Phi\bar{\Psi}\Psi \;,
\end{equation}
generates a background-dependent mass for $\Psi$.
This inflaton-induced mass changes drastically at the moment of transition from the inflationary to the hot Big-Bang cosmology: instead of (\ref{Meff1}), we now have
\begin{equation}\label{Meff2}
    M_{\rm eff}(\eta)=M(\eta)a(\eta) \;,
\end{equation}
where $M(\eta)$ can vary by many orders of magnitude during the transition, driven by the evolution of the inflaton field.
The time-varying mass (\ref{Meff2}) combines the effects due to the background geometry and due to the inflaton-fermion coupling, which, in general, cannot be disentangled.
The non-perturbative fermion production from the coupling (\ref{Int1}) during preheating has also been subject of extensive research; see, e.g.,~\cite{Giudice:1999fb,Peloso:2000hy,Nilles:2001fg} for early contributions and \cite{Garcia:2021iag} for a more recent analysis.

In this paper we revisit the non-perturbative production of (spin-$1/2$) fermions during and after preheating.
We compute analytically the Bogolyubov coefficient between the in-state inflationary modes that satisfy the Bunch--Davies vacuum condition and the out-state modes in the radiation-dominated epoch.
We account for the effects due to varying background geometry and due to possible coupling of the fermion field to the varying inflaton background.
Our main goal is to estimate analytically the sensitivity of the production mechanism to the preheating epoch.
To this end, we model the preheating by adopting an ansatz for the scale factor $a$ and the varying mass in \cref{Meff2} that interpolates smoothly between the end of inflation and the beginning of radiation-dominated stage.
We neglect the back-reaction of the produced fermions on the evolution of the metric and inflaton field.

We focus on the case of fast preheating that completes within less than the inverse inflationary Hubble rate.
Moreover, for the analytical calculation we assume the preheating to be ``instant'' \cite{Felder:1998vq}, that is, to take less than one oscillation of the (homogeneous) inflaton background around the bottom of the inflationary potential.
Such properties of the preheating are typical, e.g., for the models of Higgs inflation~\cite{Bezrukov:2007ep,Bauer:2008zj}.
Note that this setup is different from the more commonly studied situation where the non-perturbative fermion production relies on quasi-periodic oscillations of the fermion effective mass and on this mass crossing zero one or more times.

First, we study the limit of infinitely short preheating, which results in the presence of discontinuities in the derivatives of $M_{\rm eff}(\eta)$.
The discontinuity leads to the high-momentum asymptotics of the Bogolyubov coefficient being power-like, $|\beta_k|^2\sim k^{-N}$, $k\to\infty$.
We find that $N=6$ in the absence of inflaton-fermion coupling (the purely gravitational production).
Hence, the number and energy densities of produced fermions are finite, in agreement with the previous studies (see, e.g., \cite{Chung:2011ck,Herring:2020cah}).
This means that the non-perturbative gravitational production mechanism is not sensitive to the details of preheating, particle production happens during the radiation-dominated epoch, and the spectrum of produced particles is dominated by modes which were outside the horizon at the end of inflation.
On the other hand, turning on the inflaton-fermion coupling gives $N=2$ and results in the UV divergence of the number density~\cite{Parker:1971pt}.
This indicates that the mechanism is sensitive to the parameters of preheating and, in particular, to its duration.

The UV divergence is regularised by the finite duration of preheating $\tau$.
Then, the Bogolyubov coefficient falls off exponentially fast at high momenta $k\gtrsim\tau^{-1}$.
The particle spectrum is dominated by excitations which are inside the horizon at the end of inflation, with energies up to $\omega\sim\tau^{-1}$, provided that $\tau H_{dS}\ll 1$, where $H_{dS}$ is the Hubble rate at the end of inflation.
The resulting DM abundance depends on $\tau$ as well.
Interestingly, we find that in the special case $H_{dS}/M_i\ll \tau H_{dS} \ll 1$, where $M_i$ is the inflaton-induced fermion mass at the end of inflation, the particle distribution (and abundance) does not depend on $M_i$.
Moreover, with the given ansatz for $a$ and $M(\eta)$, the fermion distribution function coincides with the Fermi--Dirac one of the temperature $T_{\rm eff}=(\pi\tau)^{-1}$.
In all other cases, the distribution of produced fermions is far from the thermal one.
Note that the mechanism operates independently of the evolution of other species during preheating and, hence, is not sensitive to the preheating temperature.

As an application of our results, we consider the production of \emph{sterile neutrino} dark matter $\Psi$ during preheating in the model of Palatini Higgs inflation \cite{Rubio:2019ypq,Dux:2022kuk}, where the Higgs field plays the role of the inflaton.
The sterile neutrinos are usually coupled to the Higgs and active neutrino fields $\nu_L$ via the Yukawa interaction similar to~\eqref{Int1}.
After the electroweak symmetry breaking, this Yukawa coupling induces a mixing between the sterile and active neutrinos,
which could lead to sterile neutrino dark matter~\cite{Dodelson:1993je}.
Due to the strong constraints on the lifetime of sterile neutrino DM (see, e.g.,~\cite{Boyarsky:2018tvu}), these couplings are too small for efficient DM production.\footnote{
	We also note that the Yukawa couplings of \emph{sterile neutrino} dark matter are typically too small to give a significant contribution to the light neutrino masses (see e.g.~\cite{Boyarsky:2018tvu}).
}
In the absence of sizeable Yukawa interactions, one can also produce sterile neutrino dark matter via effective operators~\cite{Bezrukov:2011sz,Shaposhnikov:2020aen}. Here we will focus on the dimension-5 operator
\begin{equation}
\mathcal{L}_{int.} = \frac{c_5}{2\Lambda}\Phi^\dag \Phi \bar{\Psi}^C_R\Psi_R  + \text{h.c.} \;,
\end{equation}
where $\Lambda$ is the high-energy suppression scale and $C$ denotes charge conjugation.
We compare this non-perturbative production mechanism with the perturbative thermal production induced by the same operator and conclude that they are equally important for realistic preheating parameters.\:\footnote{For a different production mechanism relying on the same operator see~\cite{Anisimov:2006hv,Anisimov:2008gg}.}

In sec.~\ref{sec:setup} we derive the general expression for the Bogolyubov coefficient in terms of the in-state and out-state modes.
Sec.~\ref{sec:inst} is dedicated to particle production in the limit of infinitely short preheating, and sec.~\ref{sec:fin} discusses the finite preheating time.
In sec.~\ref{sec:HI} we review the model of Palatini Higgs inflation and use it to compare the efficiency of different production mechanisms.
Finally, in sec.~\ref{sec:concl} we summarize our main findings.
The details of the calculations are collected in several appendices.

\section{Setup}
\label{sec:setup}

Consider a massive Dirac fermion $\Psi$ with the action\:\footnote{We will comment on the case of Majorana fermion in the end of sec.~\ref{sec:setup}.}
\begin{equation}\label{S}
  S=\int\diff^4 x\sqrt{-g}\bar{\Psi}\left( i\gamma^\mu \mathcal{D}_\mu - M \right)\Psi \;,
\end{equation}
where $\mathcal{D}_\mu$ is the curved-space covariant derivative involving spin-connection.
We adopt the spatially-flat FRWL metric in the conformal time,
\begin{equation}\label{ds}
  \diff s^2=a(\eta)^2(-\diff\eta^2+\diff\vec{x}^2) \;.
\end{equation}
Substituting this metric to the action (\ref{S}) and redefining the variables as $\Psi\mapsto a^{3/2}\Psi$, $M\mapsto aM\equiv M_{\rm eff}$, we obtain the equivalent theory in the Minkowski spacetime,
\begin{equation}\label{S2}
  S=\int\diff^3\vec{x}\:\diff\eta \:\bar{\Psi} (i\gamma^\mu\partial_\mu-M_{\rm eff})\Psi \;,
\end{equation}
with the time-dependent effective fermion mass.

If one allows for the direct coupling of $\bar{\Psi}\Psi$ to the background inflaton field $\Phi$ (or the function of thereof), this coupling also contributes to the effective mass of $\Psi$.
For example, the Yukawa interaction (\ref{Int1}) leads to $M_{\rm eff}=a(M+c\Phi)$ with the redefinition $\Phi\mapsto a\Phi$.
The combination $M+c\Phi$ varies from $M_i=M+c\Phi_e$, where $\Phi_e$ is the inflaton value at the end of inflation, to $M_f=M$.
We will see that this mass variation effect can affect significantly the production rate as compared to the purely gravitational production driven by the scale factor.

From \cref{S2} the Dirac equation follows,
\begin{equation}\label{EoM}
  (i\gamma^\mu\partial_\mu-M_{\rm eff})\Psi = 0\;,
\end{equation}
and its solution in a comoving volume $V$ is written as
\begin{equation}\label{psi}
\Psi(\Vec{x}, \eta) = \frac{1}{\sqrt{V}}\sum_{\Vec{k};s=\pm 1} \e^{i \Vec{k} \cdot \Vec{x}}\left( b_{\Vec{k},s} \UU_{k,s}(\eta) + d^{\dagger}_{-\Vec{k},s}\VV_{-k,s}(\eta) \right) \; ,
\end{equation}
where $\UU_{k,s}$, $\VV_{k,s}$ are the four-component orthogonal and normalised spinors,
\begin{equation}\label{NormSpinors}
    \UU^\dag_{k,s}(\eta)\UU_{k,s'}(\eta)=\delta_{s,s'} \;, ~~~ \VV^\dag_{-k,s}(\eta)\VV_{-k,s'}(\eta)=\delta_{s,s'} \;, ~~~
    \UU^\dag_{k,s}(\eta)\VV_{-k,s'}(\eta)=0 \;.
\end{equation}
It is convenient to reduce \cref{EoM} to a pair of equations for complex scalar mode functions.
To this end, one writes (see, e.g., \cite{Herring:2020cah})
\begin{subequations}\label{UUVV}
\begin{align}
       \UU_{k,s}  & = \left( i\gamma^0\partial_\eta-\vec{\gamma}\cdot\vec{k}+M_{\rm eff}(\eta)\right)f_k(\eta)u_s \;, \label{UU} \\
       \VV_{-k,s} & = \left( i\gamma^0\partial_\eta-\vec{\gamma}\cdot\vec{k}+M_{\rm eff}(\eta)\right)g_k(\eta)v_s \;. \label{VV}
\end{align}
\end{subequations}
We use the Dirac representation for the gamma matrices.
The constant spinors $u_s$, $v_s$ can be chosen as $u_s=(\xi_s,0)^T$, $v_s=(0,\xi_s)^T$, where $\xi_s$ are the helicity eigenstates obeying $\vec{\sigma}\cdot \vec{k}\:\xi_s=sk\xi_s$, $s=\pm 1$.
From \cref{EoM,psi,UUVV} we deduce the equations for the mode functions
\begin{subequations}\label{Eq_fg}
\begin{align}
    & f_k''+(k^2+M_{\rm eff}^2-iM'_{\rm eff})f_k=0 \;, \label{Eq_f} \\
    & g_k''+(k^2+M_{\rm eff}^2+iM'_{\rm eff})g_k=0 \;, \label{Eq_g}
\end{align}
\end{subequations}
where prime means the derivative with respect to $\eta$.

Let us discuss the boundary conditions for the mode functions.
Consider the inflationary epoch, with the Hubble rate $H_{dS}=a'_{dS}/a^2_{dS}$.
We neglect slow-roll corrections to the background de Sitter geometry.
Then $H_{dS}$ is constant, and we have
\begin{equation}\label{a_dS}
  a_{dS} = \frac{1}{H_{dS}( \eta_R - \eta)}
\end{equation}
with $\eta_R$ a constant.
Substituting this to \cref{Meff2,Eq_fg} we obtain
\begin{equation}\label{Eq_Infl}
  f_k''+ \left( k^2 - \frac{\nu^2-1/4}{( \eta_R-\eta)^2} \right)f_k = 0 \;,
\end{equation}
where $\nu = 1/2+i M_i/H_{dS}$, and the analogous equation for $g_k$.
We require the in-state modes $f_{\text{in},k}$, $g_{\text{in},k}$ to be in the Bunch--Davies vacuum.
This implies
\begin{equation}\label{BD}
  f_{\text{in},k}\to\e^{-ik\eta } \;, ~~~ g_{\text{in},k}\to f_{\text{in},k}^* \;, ~~~ \eta\to-\infty
\end{equation}
for modes which are deep inside the Hubble radius during inflation.
From here and \cref{Eq_fg} we conclude that $g_{\text{in},k}=f_{\text{in},k}^*$, and the in-state spinors take the form
\begin{equation}\label{UV_in}
    \UU_{k,s}^{\rm in} =\mathcal{N}_{\rm in}\begin{pmatrix} \xi_s F_{\text{in},k}\\ \xi_s k s f_{\text{in},k} \end{pmatrix} \; , \qquad  \VV_{-k,s}^{\rm in}= \mathcal{N}_{\rm in}\begin{pmatrix} -\xi_s k s f^{*}_{\text{in},k}\\ \xi_s F^{*}_{\text{in},k} \end{pmatrix} \; ,
\end{equation}
where $F_{\text{in},k}=if'_{\text{in},k}+M_{\rm eff}f_{\text{in},k}$ and $\mathcal{N}_{\rm in}$ is the constant normalisation factor.
The solution of \cref{Eq_Infl} that satisfies the vacuum condition (\ref{BD}) is
\begin{equation}\label{f}
 f_{\text{in},k} = \sqrt{\frac{\pi k (\eta_R-\eta)}{2}}\e^{\frac{i\pi}{2}\left(\nu+\frac{1}{2}\right)} H_{\nu}^{(1)}(k(\eta_R-\eta)) \;,
\end{equation}
where $H_{\nu}^{(1)}(x)$ is the Hankel function of the first kind.

Consider now the radiation-dominated epoch.
The Hubble rate is given by $H_{rd}=H_R/a^2_{rd}$ with $H_R=1.66\sqrt{g_{*,eq}}T_{eq}^2a_{eq}^2/M_P\approx 10^{-44}$ GeV.
Here $g_{*,eq}=3.36$, $T_{eq}\approx 0.7$ eV, $a_{eq}\approx 3\cdot 10^{-4}$ are the effective number of relativistic degrees of freedom, the temperature and the scale factor at the moment of matter-radiation equality, respectively, and $M_P=2.4\cdot 10^{18}$ GeV is the reduced Planck mass.
Using $H_{rd}=a'_{rd}/a^2_{rd}$ we find
\begin{equation}\label{a_rd}
  a_{rd}=H_R ( \eta_R + \eta) \;.
\end{equation}
The constant $\eta_R$ is found from matching \cref{a_dS,a_rd} at the transition from inflation to radiation-dominated epoch.
Let the transition happen at $\eta=\eta_e=0$.
Then, from the continuity of the scale factor, $a_{dS}(0)=a_{rd}(0)$, it follows that
\begin{equation}\label{etaR}
  \eta_R=\frac{1 }{\sqrt{H_{dS}H_R}} \;.
\end{equation}
In what follows we assume that the transition period---the preheating---is complete within the (conformal) time $\tau_\eta \ll \eta_R$.
We will see that this is a reasonable assumption when discussing the fast preheating.
In particular, it is satisfied in the model studied in sec.~\ref{sec:HI}.
Substituting \cref{Meff2,a_rd} to \cref{Eq_f,Eq_fg} we obtain at the radiation-dominated epoch
\begin{equation}\label{Eq_RD}
  f_k'' + (k^2 + M_f^2 H_R^2 (\eta_R+\eta)^2-iM_fH_R)f_k =0 \; ,
\end{equation}
and similarly for $g_k$.
The out-state vacuum is defined at a time $\eta=\eta_*(k)$ at which the gravitational production of the mode with comoving momentum $k$ ceases.
The vacuum modes $f_{\text{out},k}$, $g_{\text{out},k}$ satisfy the boundary condition
\begin{equation}\label{RD}
 f_{\text{out},k} \rightarrow \e^{-i \int^{\eta} \diff \eta' \omega_{\eta',k}} \;, ~~~ g_{\text{out},k}\to f^*_{\text{out},k} \;, ~~~ \eta\gtrsim\eta_* \;, ~~~  \omega_{\eta,k}^2 = k^2 + M_f^2a_{rd}^2 \; .
\end{equation}
It follows that $g_{\text{out},k}=f_{\text{out},k}^*$ and the out-state spinors take the form
\begin{equation}\label{UV_out}
    \UU_{k,s}^{\rm out} =\mathcal{N}_{\rm out}\begin{pmatrix} \xi_s F_{\text{out},k}\\ \xi_s k s f_{\text{out},k} \end{pmatrix} \; , \qquad  \VV_{-k,s}^{\rm out}= \mathcal{N}_{\rm out}\begin{pmatrix} -\xi_s k s f^{*}_{\text{out},k}\\ \xi_s F^{*}_{\text{out},k} \end{pmatrix}
\end{equation}
with $F_{\text{out},k}=if'_{\text{out},k}+M_{\rm eff}f_{\text{out},k}$ and $\mathcal{N}_{\rm out}$ the normalisation factor.
The solution of \cref{Eq_RD} satisfying \cref{RD} reads as
\begin{equation}\label{f_RD}
 f_{\text{out},k}(\eta) = D_\alpha(\sqrt{2} \e^{\frac{i\pi}{4}}\sqrt{M_f H_R} (\eta_R+\eta)) \; ,
\end{equation}
where $\alpha = -1 - i k^2/(2 M_f H_R)$ and $D_{\alpha}$ is the parabolic cylinder function.

We take the solution $\bar{f}_k$ of \cref{Eq_f}, which coincides with the in-state solution $f_{\text{in},k}$ at $\eta<0$, $|\eta|\gg\tau_\eta$, and continue it to the region $\eta>0$.
Introduce the spinors $\bar{\UU}_{k,s}$, $\bar{\VV}_{-k,s}$ associated with the modes $\bar{f}_k$.
At $\eta=\eta_*$ we expand
\begin{equation}\label{Expansion}
  \bar{\UU}_{k,s}=\alpha_{k,s}\:\UU^{\text{out}}_{k,s}+\beta_{k,s}\:\VV^{\text{out}}_{-k,s} \;,
\end{equation}
and similarly for $\bar{\VV}_{-k,s}$.
Using the normalisation and orthogonality conditions (\ref{NormSpinors}), we find the Bogolyubov coefficient
\begin{equation}\label{Beta}
  \beta_{k,s}=\VV^{\text{out}\dag}_{-k,s}\:\bar{\UU}_{k,s} \;.
\end{equation}
Multiplying this by the complex conjugate and using \cref{NormSpinors,UV_out,BD}, we arrive at the following expression
\begin{equation}\label{BetaMain}
     |\beta_k|^2 = \frac{1}{2} |\bar{f}_k|^2\frac{\left\vert  \bar{f}_k'/\bar{f}_k-f_{\text{out},k}'/f_{\text{out},k}-i(M_\text{in}-M_\text{out}) \right\vert^2}{k^2+\left\vert f_{\text{out},k}'/f_{\text{out},k}-i M_{\text{out}} \right\vert^2 } \;,
\end{equation}
where we omitted the spin index, and where
$M_\text{in} = M_\text{eff}(\eta_*-\delta)$ and
$M_\text{out} = M_\text{eff}(\eta_*+\delta)$ with $\delta \rightarrow 0$.
Note that the term $\propto M_\text{out} - M_\text{in}$ vanishes if $M_\text{eff}$ is continuous, but is finite when $M_\text{eff}$ has a discontinuity (as in the limit $\tau_\eta \rightarrow 0$ and $\eta_* \rightarrow 0$).

The derivation of the Bogolyubov coefficient (\ref{BetaMain}) goes unchanged for the Majorana fermion $\Psi$.
This is because the vacuum conditions (\ref{BD}), (\ref{RD}) are compatible with the reality condition for the Majorana field.
As a result, the spinors (\ref{UV_in}), (\ref{UV_out}) satisfy the charge conjugation relation
\begin{equation}
    i\gamma^2\left( \UU^{\rm in}_{k,\lambda} \right)^* = \VV^{\rm in }_{k,\lambda} \;, ~~~ i\gamma^2 \left( \VV^{\rm in}_{k,\lambda} \right)^* = \UU^{\rm in }_{k,\lambda} \;, ~~~ \lambda=1,2
\end{equation}
and similartly for $\UU^{\rm out}_{k,\lambda}$, $\VV^{\rm out}_{k,\lambda}$.

The comoving particle number $n$ and energy $\rho$ densities of produced Dirac fermions (particles and antiparticles) are evaluated at $\eta>\eta_*$ as follows
\begin{align}
 & n=\frac{2 }{\pi^2a^3}\int_0^\infty\diff kk^2|\beta_k|^2 \;, \label{n} \\
 & \rho=\frac{2 }{\pi^2 a^4}\int_0^\infty\diff k k^2\omega_{\eta,k} |\beta_k|^2 \label{rho} \;.
\end{align}
If $\Psi$ is of Majorana nature, one should divide by $2$ in these expressions, since in this case the particle and antiparticle are identical.

Finally, the abundance $\O_\Psi$ is conveniently evaluated at the moment of matter-radiation equality.
Normalising to the DM abundance $\O_{DM}$, we find
\begin{equation}\label{O1}
    \frac{\O_\Psi}{\O_{DM}}=\frac{a_{eq}^3\rho }{\O_{DM}\rho_c} \;,
\end{equation}
where $\rho_c=0.4\cdot 10^{-46}$ GeV is the critical energy density of the universe.

\section{The limit of infinitely short preheating}
\label{sec:inst}

We first compute the distribution function $|\beta_k|^2$ in the limit of infinitely short preheating, $\tau_\eta\to 0$.
We are interested in the behavior of $|\beta_k|^2$ at various momentum scales and, in particular, in its high-energy asymptotics.
Depending on the value of $N$ in the asymptotics $|\beta_k|^2\sim k^{-N}$ at $k\to\infty$, the particle number or energy density of produced fermions may or may not UV diverge.
In the former case, the divergence is regularised by invoking the finite preheating time.
The particle spectrum is then dominated by energetic modes, and the resulting abundance depends on $\tau_\eta$.
In the latter case, the spectrum is dominated by soft modes and there is no sensitivity to preheating.

We start with assuming no coupling between the fermion and inflaton fields.
In our notation this means $M_i=M_f\equiv M$, and the particle production is only due to the varying background geometry.
In this case, we are mainly interested in the regime $M\ll H_{dS}$.
Otherwise, the production during or after preheating is suppressed.\:\footnote{The situation can be different in the case when the inflaton field undergoes rapid oscillations during preheating \cite{Ema:2019yrd}.
In general, the abundant production of ``superheavy'' (bosonic or fermionic) DM with $M\gtrsim H_{dS}$ in the early universe is possible in various scenarios \cite{Kuzmin:1998uv,Chung:1998zb,Kuzmin:1998kk,Chung:1998bt,Chung:2001cb,Ema:2015dka,Ema:2016hlw,Ema:2018ucl, Hashiba:2018iff,Chung:2018ayg,Ema:2019yrd,Babichev:2020yeo}.}

From the general expression (\ref{BetaMain}) we can discern two characteristic momentum scales,
\begin{equation}\label{k1k2}
    k_1=\sqrt{M H_R} \;, ~~~ k_2=\sqrt{H_{dS}H_R} \;.
\end{equation}
Let us discuss their physical meaning.
The modes with $k\gtrsim k_1$ are relativistic at the moment $\eta_1$ at which $H(\eta_1)=M$, and the modes with $k\ll k_1$ are non-relativistic at this moment.
Indeed, for the physical momentum at $\eta=\eta_1$ we find
\begin{equation}
    \frac{k_1 }{a_1}=k_1\frac{1 }{a_e}\frac{a_e}{a_1}=\sqrt{MH_R}\sqrt{\frac{H_{dS}}{H_R}}\sqrt{\frac{M}{H_{dS}}}=M \;.
\end{equation}
Hence, for $k<k_1$ we expect $|\beta_k|^2\approx 1/2$, while for $k>k_1$ a power suppression is expected \cite{Chung:2011ck,Herring:2020cah}.
Next, the modes with $k>k_2$ are inside the horizon at the end of inflation, and the modes with $k<k_2$ are outside the horizon at this moment.
Indeed,
\begin{equation}
    \frac{k_2 }{a_e}=\sqrt{H_{dS}H_R}\sqrt{\frac{H_{dS}}{H_R}}=H_{dS} \;.
\end{equation}
Hence, we expect the particle production with $k>k_2$ be further suppressed \cite{Chung:2011ck}.
Note that the condition $M\ll H_{dS}$ implies $k_1\ll k_2$.

\begin{figure}[t]
    \center{
        \begin{minipage}[h]{0.45\linewidth}
            \center{\includegraphics[width=0.99\linewidth]{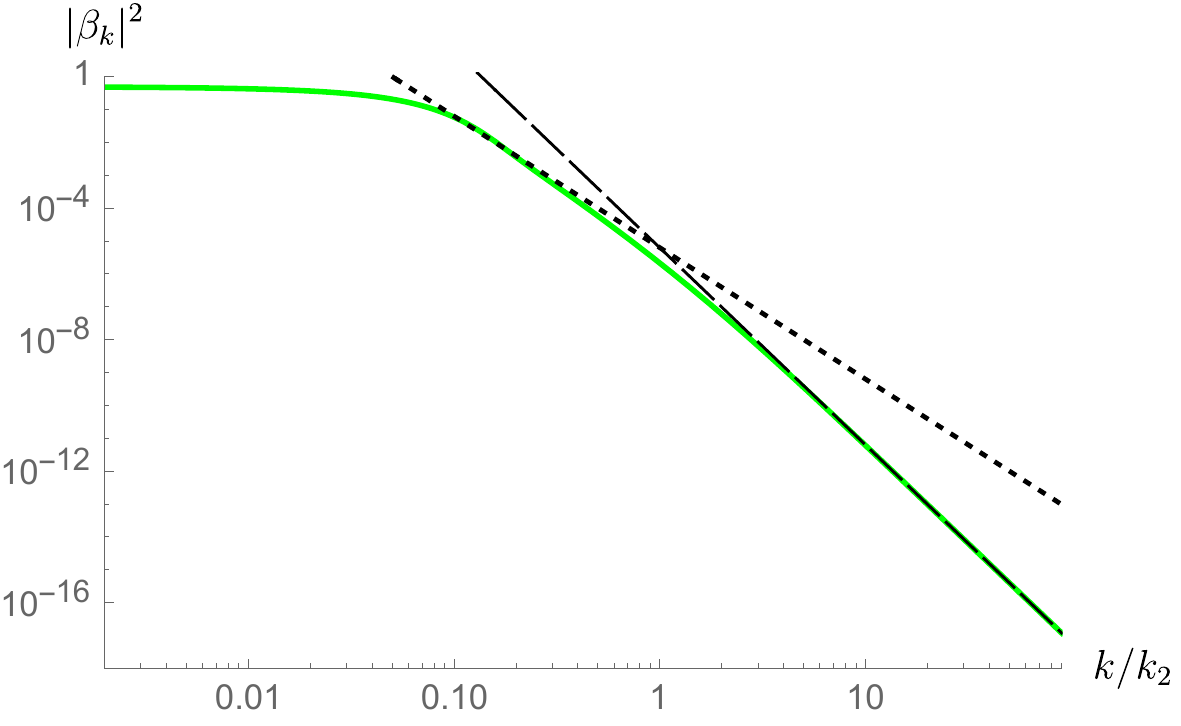}}
        \end{minipage}
        \hfill
        \begin{minipage}[h]{0.45\linewidth}
            \center{\includegraphics[width=0.99\linewidth]{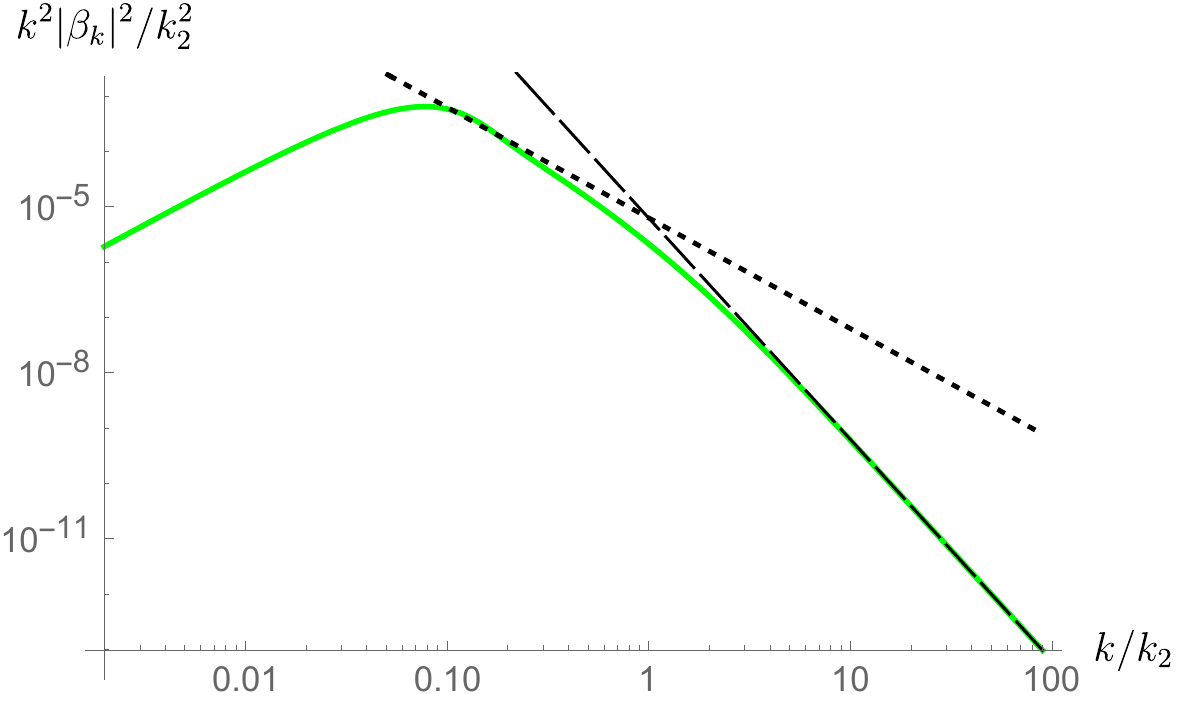}}
        \end{minipage}
    }
\caption{The distribution function \textbf{(left)} and spectral density \textbf{(right)} of particles in the limit of infinitely short preheating and no inflaton-fermion coupling.
We take $M=10^8$ GeV and $H_{dS}=10^{10}$ GeV, so that $k_1=0.1k_2$ (see \cref{k1k2}).
The dotted and dashed lines show the second and third asymptotics in \cref{BetaInstSmallM}, respectively.
}
\label{fig:InstBeta}
\end{figure}

To confirm these qualitative arguments, we calculate the Bogolyubov coefficient analytically in three different regimes.
We leave the details of the calculation to Appendix~\ref{App:tau=0} and present the result:
\begin{equation}
    \label{BetaInstSmallM}
    |\beta_k|^2_{k\lesssim k_1} = \frac{1 }{2 } \;, ~~~ |\beta_k|^2_{k_1\lesssim k\lesssim k_2} = \frac{k_1^4}{16k^4} \;, ~~~ |\beta_k|^2_{k_2\lesssim k} = \frac{k_1^4k_2^2}{16k^6} \;, ~~~ k_1\ll k_2 \;.
\end{equation}
These asymptotics agree with the more qualitative results of Ref.~\cite{Chung:2011ck}.
\footnote{Note that the large momentum limit in \cref{BetaInstSmallM} also holds for conformally coupled scalars~\cite{Chung:1998zb}.}
Next, we compute $|\beta_k|^2$ numerically using \cref{f,f_RD,BetaMain}.
We plot it in Fig.~\ref{fig:InstBeta} for the particular values of $M$ and $H_{dS}$.
We see that the asymptotics (\ref{BetaInstSmallM}) correctly reproduce the behavior of $|\beta_k|^2$ in the respective ranges of $k$.
Notice that the particle distribution shown in Fig.~\ref{fig:InstBeta} cannot be approximated by the thermal distribution function: the particle spectrum is far from the equilibrium one.

The large-energy behavior of $|\beta_k|^2$ is such that the integrals in \cref{n,rho} are convergent.
This is despite the presence of discontinuity in the derivative of the scale factor and the effective fermion mass.
It follows that the particle spectrum is dominated by the modes with $k\sim k_1$ (see Fig.~\ref{fig:InstBeta}), while the contribution from the modes with $k\gtrsim k_2$ is suppressed.
We conclude that the mechanism of gravitational production of fermions is not sensitive to preheating.\:\footnote{The same conclusion was reached in Ref.~\cite{Herring:2020cah}.}

\begin{figure}[t]
\centering
  \includegraphics[width=0.5\linewidth]{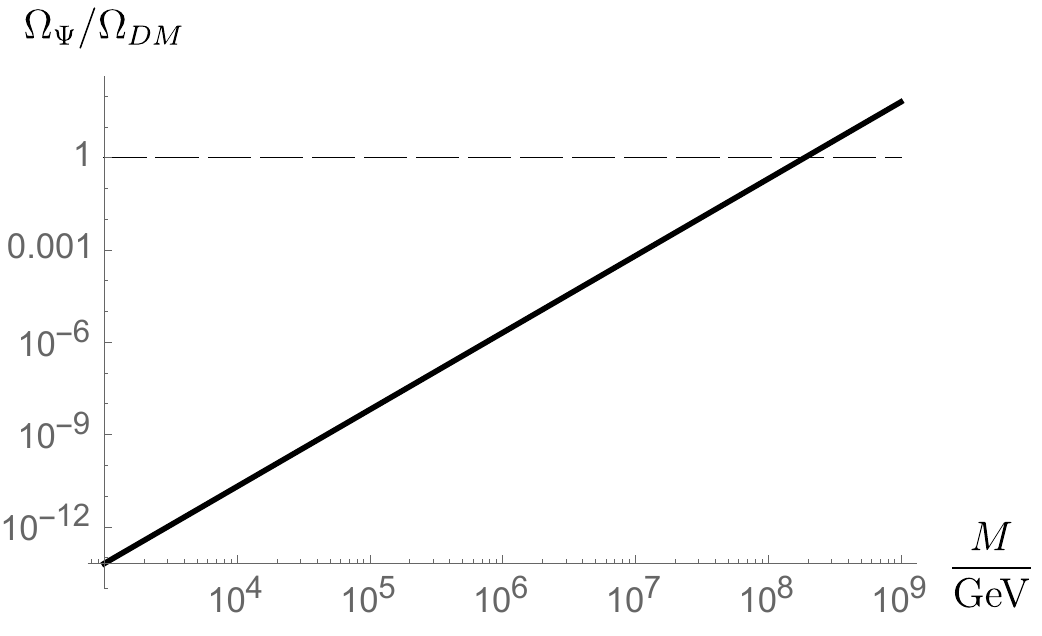}
\caption{Particle abundance at different fermion mass $M\ll H_{dS}$, assuming no inflaton-fermion coupling.
The dashed line corresponds to $\Omega_\Psi/\Omega_{DM}=1$.}
\label{fig:InstOmega}
\end{figure}

Let us comment on ``superheavy'' particles with $M\gtrsim H_{dS}$.
One can show that in this case $|\beta_k|^2$ is bounded from above by the ratio $(H_{dS}/M)^5$ with the maximum attained at $k\sim k_1$, see Appendix~\ref{App:tau=0} for details.\:\footnote{This agrees with the results of Ref.~\cite{Chung:2011ck}.}
Hence, as expected, the gravitational production of such particles after inflation is suppressed.

Using \cref{n,rho} we compute the number and energy densities of the produced particles (in the regime $M\ll H_{dS}$),
\begin{equation}
    n= \frac{0.2 M^{3/2}H_R^{3/2} }{ \pi^2 a^3} \;, ~~~ \rho= Mn \;.
\end{equation}
The coefficient is found by integrating numerically \cref{n}.
The corrections to these expressions are of order $k_1/k_2=\sqrt{M/H_{dS}}$ and $k_2/(aM)$.
The first is small by assumption.
The second is tiny at $a=a_{eq}$ in view of the bound $H_{dS}\lesssim 10^{14}$ GeV, which follows from the constraint on the tensor-to-scalar ratio \cite{Planck:2018jri}, and which gives $k_2/a_{eq}\lesssim 10^{-3}$ eV, much below the allowed mass of the fermion DM candidate \cite{Tremaine:1979we}.
Finally, from \cref{O1} we obtain the dark fermion abundance $\O_\Psi$ at the moment of matter-radiation equality,
\begin{equation}
    \frac{\Omega_\Psi }{\Omega_{DM}}=\left(\frac{M }{1.9\cdot 10^8\:\rm{GeV}}\right)^{5/2} \;,
\end{equation}
which is consistent with the previous studies on the gravitational production of fermionic DM~\cite{Kuzmin:1998uv,Kuzmin:1998kk,Chung:2011ck,Herring:2020cah}.
We plot the abundance in Fig.~\ref{fig:InstOmega} as a function of fermion mass.
Note that it does not depend on $H_{dS}$ (as long as $M\ll H_{dS}$) manifesting that the production happens during the radiation-dominated epoch.

\begin{figure}[t]
    \centering
    \includegraphics[width=0.65\linewidth]{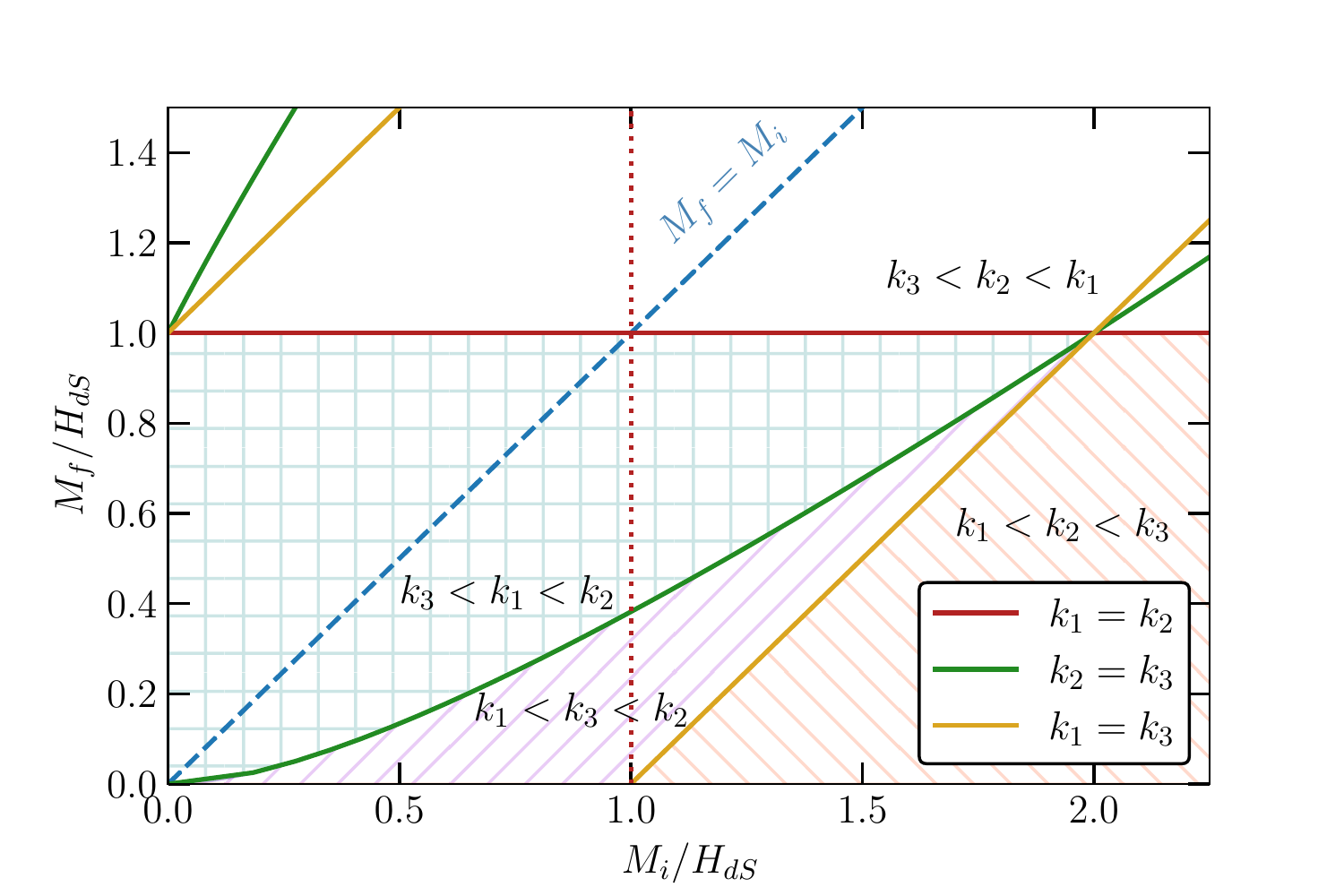}
    \caption{
    The different regimes of DM production based on the relations between the three momentum scales $k_1$, $k_2$ and $k_3$.
    The possible cases are determined by the ratios of the masses $M_{i,f}$ to the Hubble rate $H_{dS}$.
    }
    \label{fig:Regimes}
\end{figure}

We now turn to a more general situation and allow the field $\Psi$ to interact with the inflaton field so that $M_i\neq M_f$.
We are agnostic of the particular form of this interaction; as an example, one can take the Yukawa coupling (\ref{Int1}).
It is natural to expect that the mass $|M_i|$ induced by the inflaton background is much larger than the bare fermion mass $M_f$.
Hence, we expect $M_f$ to enter the answer only insofar as the (nonrelativistic) fermion energy density is proportional to it.
For this reason, the sign of the coupling constant in \cref{Int1} is not important, and, for concreteness, we take $c>0$, hence $M_i\gtrsim M_f$.
Furthermore, we assume that $M_f\ll H_{dS}$.
This is a reasonable assumption, since otherwise the particle production after inflation is suppressed, as in the constant-mass case.\:\footnote{As discussed in introduction, this reasoning is valid in the regime of fast and instant preheating, which includes the limit of infinitely short preheating.}
Note that the inflaton-induced mass $M_i$ does not have to be below $H_{dS}$ for the production mechanism to be efficient, as we will see shortly.

By definition of scale $k_1$, one should now take
\begin{equation}\label{k1new}
    k_1=\sqrt{M_f H_R} \;.
\end{equation}
The scale $k_2$ remains unchanged.
Furthermore, from \cref{BetaMain} we notice the presence of yet another scale
\begin{equation}\label{k3}
    k_3=(M_i-M_f)\sqrt{\frac{H_R }{H_{dS}}} \;.
\end{equation}
Using \cref{k1k2,k1new,k3} and the condition $M_f\ll H_{dS}$, we obtain three regimes:\\
$(i) ~~~~ k_2>k_1>k_3$. This corresponds to $M_i< H_{dS}$ and $M_i^2< M_fH_{dS}$.\\
$(ii) ~~~ k_2>k_3>k_1$. This happens when $M_i< H_{dS}$ but $M_i^2> M_fH_{dS}$.\\
$(iii) ~~ k_3>k_2>k_1$. This happens when $M_i>H_{dS}$.
In this case $k_3\approx k_2 M_i/H_{dS}$.\\
The limit of vanishing fermion-inflaton coupling belongs to regime $(i)$.
Increasing the coupling brings us first to case $(ii)$ and, eventually, to case $(iii)$.
These relations between the momentum scales are illustrated in Fig.~\ref{fig:Regimes}.
Note that even in case $(iii)$ of initially superheavy fermion, the associated fermion-inflaton coupling need not be large.
For example, taking the interaction (\ref{Int1}) and the typical inflaton value at the end of inflation $\Phi_e\sim M_P$, the condition $M_i>H_{dS}$ implies $c>H_{dS}/M_P$.

\begin{figure}[t]
\centering
  \includegraphics[width=0.5\linewidth]{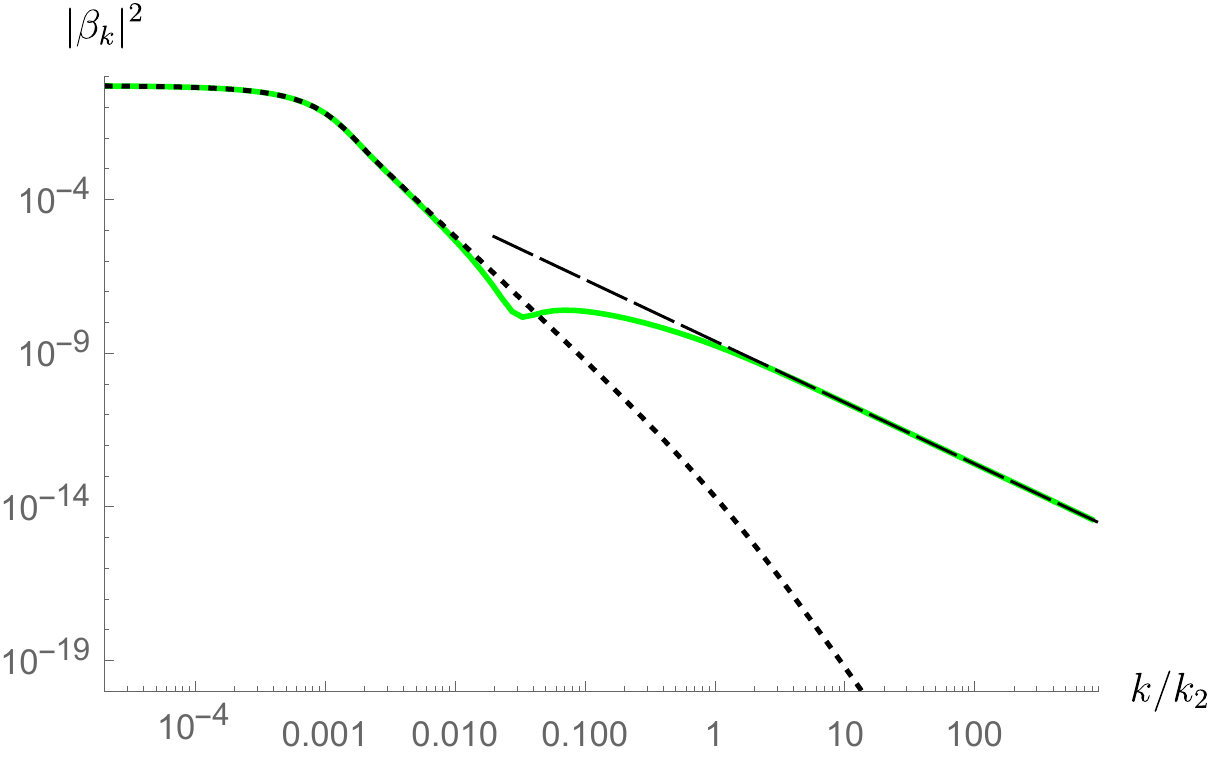}
\caption{Bogolyubov coefficient in the limit of infinitely short preheating and varying fermion mass.
The comoving momentum $k_2$ is defined in \cref{k1k2}.
We take $M_i=10^6$ GeV, $M_f=10^4$ GeV, and $H_{dS}=10^{10}$ GeV.
The dotted line shows $|\beta_k|^2$ for the same $M_f$, $H_{dS}$, and $M_i=M_f$.
The dashed line represents the asymptotics (\ref{BetaInstVarMassHighK}).
We see that $|\beta_k|^2$ has a quadratic pole at infinity.
}
\label{fig:InstBetaM1}
\end{figure}

Irrespective of the particular ordering of the scales, the high-momentum asymptotics of the Bogolyubov coefficient is the same.
We delegate the calculation to Appendix \ref{App:tau=0} and give the result:
\begin{equation}\label{BetaInstVarMassHighK}
  |\beta_k|^2_{k_2^*\lesssim k} = \frac{k_3^2 }{4 k^2} \;,
\end{equation}
where
\begin{equation}\label{k2*}
    k_2^*=\max( k_2\:, k_3) \;.
\end{equation}
Crucially, the time-variation of the fermion mass changes the order of singularity in $|\beta_k|^2$ at large $k$ from $N=6$ to $N=2$.
This is shown in Fig.~\ref{fig:InstBetaM1}, where we evaluate numerically the expression (\ref{BetaMain}) for particular values of $M_i$, $M_f$, $H_{dS}$, and compare it with the constant-mass case $M_i=M_f$.
In fact, the quadratic pole can be seen directly from the last term in \cref{BetaMain}.
Because of it, the total number and energy densities are UV-divergent.
The divergence indicates that the approximation of infinitely short preheating is not applicable to the case of varying fermion mass.
Physically, this means that the particle production is dominated by the modes which are inside the horizon at the end of inflation and which are sensitive to the parameters of preheating.

\begin{figure}[t]
\centering
  \includegraphics[width=0.65\linewidth]{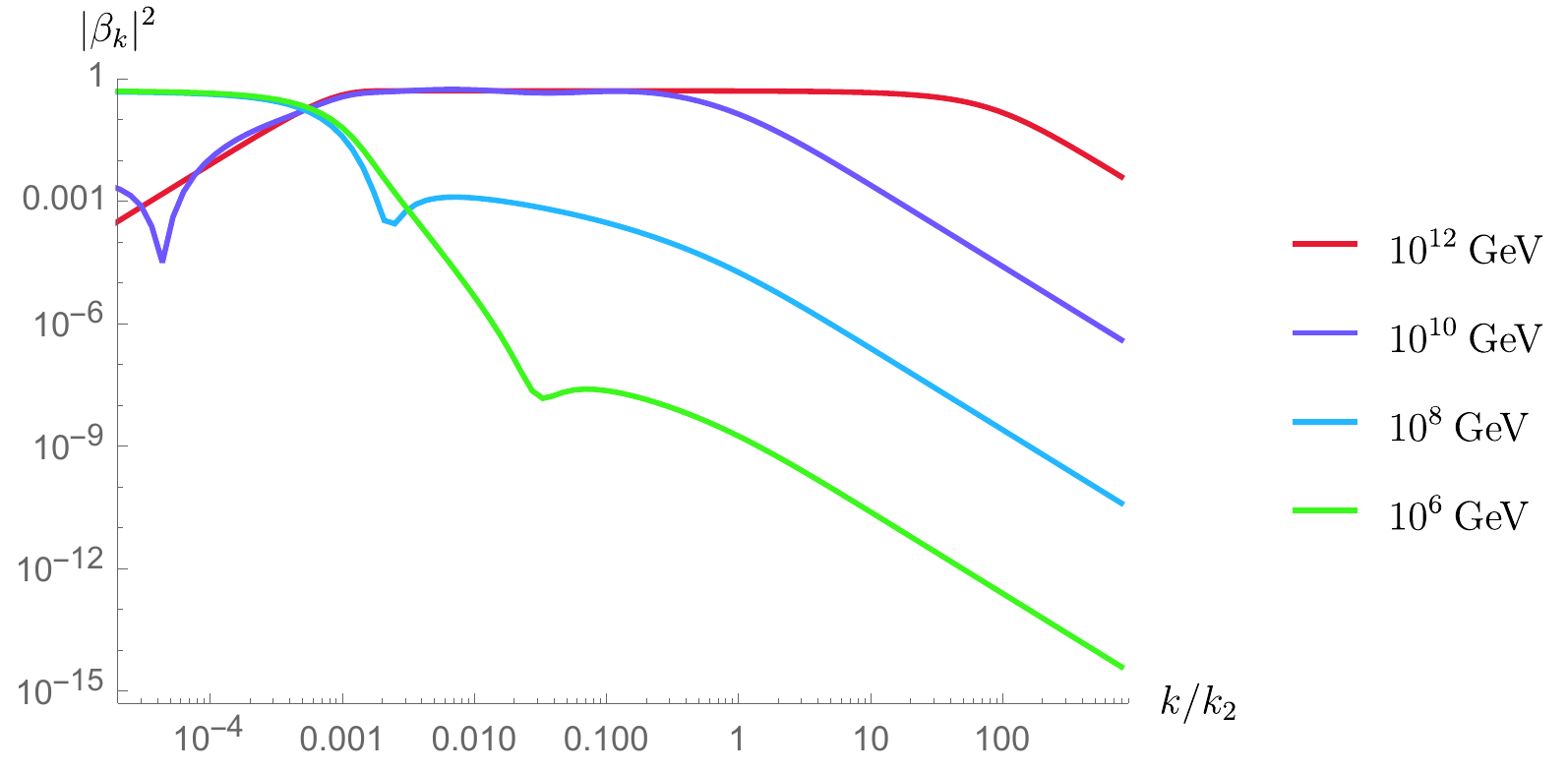}
\caption{Bogolyubov coefficient in the limit of infinitely short preheating and varying fermion mass.
We take $M_f=10^4$ GeV, $H_{dS}=10^{10}$ GeV, and several values of $M_i$.
When $M_i\gtrsim H_{dS}$, we see that $|\beta_k|^2\approx 1/2$ in the range $k_1\lesssim k\lesssim  k_3\approx k_2 M_i/H_{dS}$ (see \cref{k1k2,k3}).
}
\label{fig:InstBetaM2}
\end{figure}

Regularisation of the Bogolyubov coefficient by the finite preheating time does not affect long-wavelength modes.
Hence, assuming that $\tau_\eta^{-1}\gg k_2^*$, it is still useful to investigate the behavior of $|\beta_k|^2$ at lower momentum scales.
In cases $(i)$ and $(ii)$, the shape of $|\beta_k|^2$ at $k\ll k_2$ is the same as in the constant-mass case, \cref{BetaInstSmallM}.
This follows from the analytical calculation presented in Appendix~\ref{App:tau=0} and is confirmed by the numerical calculation, see Fig.~\ref{fig:InstBetaM1} for the illustration.
The things are qualitatively different in case $(iii)$ of superheavy inflaton-induced mass.
Examination of various approximations of the functions in \cref{f,f_RD} leads to\:\footnote{The behavior changes once again at exponentially small values of $k$, but this has no observable consequences, see Appendix~\ref{App:tau=0} for details.
}
\begin{equation}\label{BetaInstVarMSmallK}
    |\beta_k|^2_{k\lesssim k_1}=\frac{\pi k^2 }{4k_1^2} \;, ~~~ |\beta_k|^2_{k_1\lesssim k\lesssim k_2}=\frac{1 }{2}  \;, ~~~ k_1\ll k_2\lesssim k_3 \;.
\end{equation}
We see numerically that, in fact, the constant value persists until $k\sim k_3$ at which it is followed by the power asymptotics (\ref{BetaInstVarMassHighK}).
This is demonstrated in Fig.~\ref{fig:InstBetaM2} where we plot $|\beta_k|^2$ at fixed $M_f$, $H_{dS}$ and for different values of $M_i$ spanning all three regimes $(i)$---$(iii)$.

Thus, energetic particles can be produced copiously during preheating by the non-perturbative gravitational mechanism once it is assisted by the time-varying fermion mass.
Moreover, the mechanism works efficiently even for initially superheavy fermions, provided that their effective mass drops together with the Hubble rate.

\section{Fast preheating}
\label{sec:fin}

In the previous section we concluded that the non-perturbative gravitational mechanism of fermionic DM production works efficiently during (the sufficiently short) preheating if the direct coupling of the DM to the inflaton background field is allowed.
We would like to evaluate analytically how the properties of particles created this way depend on the parameters of preheating.
In an attempt to limit the model-dependence as much as possible, we adopt a simple ansatz for the scale factor $a$ and the time-varying fermion mass $M$ which interpolate between their values at the end of inflation and the beginning of radiation-dominated epoch:
\begin{equation}\label{a_tau}
a =  \frac{a_{dS}}{2} \left(1 - \th\frac{\eta}{\tau_\eta}\right) + \frac{a_{rd}}{2} \left(1 + \th\frac{\eta}{\tau_\eta}\right) \; ,
\end{equation}
and
\begin{equation}\label{M_tau}
    M=\frac{M_i+M_f}{2 }+\frac{M_f-M_i}{2 }\th\frac{\eta }{\tau_\eta} \;.
\end{equation}
Both interpolations are controlled by a single parameter $\tau_\eta$ which sets the (conformal) preheating time.
This way we smear the discontinuity in the derivatives of $a$ and $M$ and regularise the quadratic pole of $|\beta_k|^2$ at large $k$.
Speaking physically, \cref{M_tau} corresponds to the instant preheating which completes within half-oscillation of the inflaton field around the bottom of the inflationary potential~\cite{Felder:1998vq}.
Furthermore, we require the preheating to be fast in the sense that
\begin{equation}\label{tau}
    \tau_\eta \ll \eta_R = k_2^{-1} \;.
\end{equation}
This condition ensures that the two regimes of DM production---the one dominated by the long wavelength modes ($k\ll k_2$) produced after preheating and the one dominated by the short wavelength modes ($k\gtrsim k_2$) produced at the moment of preheating---are well separated and can be treated independently.
Note also that, in case $(iii)$, \cref{tau} is compatible with both $\tau_\eta^{-1}\gg k_3$ and $\tau_\eta^{-1}\lesssim k_3$, where $k_3$ is given in \cref{k3}.
We will see that these two possibilities lead to qualitatively different predictions for the DM spectrum and abundance.
Finally, it is useful to write \cref{tau} in terms of the physical preheating time $\tau$.
Using \cref{a_tau,k1k2} we obtain $\tau=\tau_\eta\sqrt{H_R/H_{dS}}$ and
\begin{equation}\label{PhysTau}
    \tau H_{dS} \ll 1 \;,
\end{equation}
i.e. the preheating must be complete within less than the inflationary Hubble time.
In sec.~\ref{sec:HI} we will see that such fast preheating can indeed be realised in a realistic model of inflation.

\begin{figure}[t]
	\center{
		\begin{minipage}[h]{0.45\linewidth}
			\center{\includegraphics[width=0.99\linewidth]{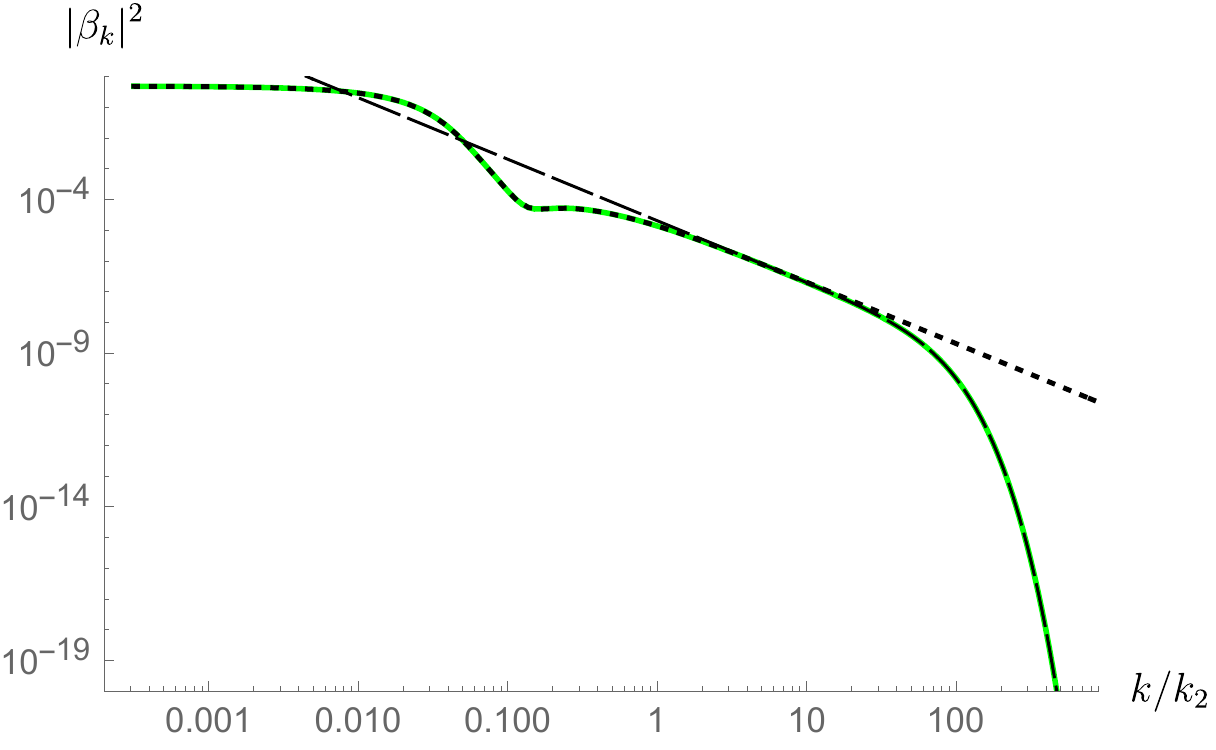}}
		\end{minipage}
		\hfill
		\begin{minipage}[h]{0.45\linewidth}
			\center{\includegraphics[width=0.99\linewidth]{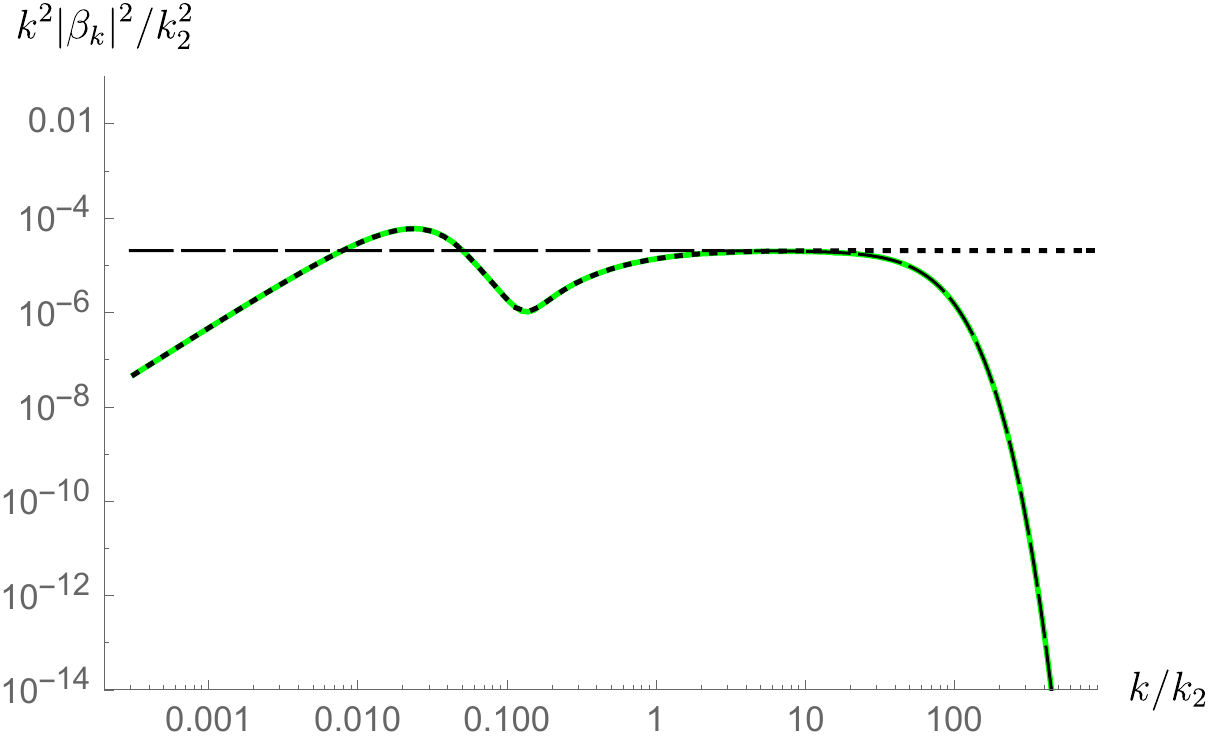}}
		\end{minipage}
	}
	\caption{\textbf{Left:} The particle distribution function with a finite preheating time.
		We take $M_i=10^8$ GeV, $M_f=10^7$ GeV, $H_{dS}=10^{10}$ GeV, and $\tau_\eta^{-1}=10^2k_2$.
		The dotted and the dashed lines represent the limit $\tau_\eta\to 0$ and the analytic expression (\ref{BetaTauLargeK}), respectively.
		\textbf{Right:} The particle spectral density at the same parameters.
		We see that the spectrum is dominated by energetic modes with $k_2\lesssim k\lesssim\tau_\eta^{-1}$.
		The scale $k_2$ is defined in \cref{k1k2}.
	}
	\label{fig:TauBeta1}
\end{figure}

Using \cref{a_tau,M_tau,tau}, one can approximately write the effective fermion mass (\ref{Meff2}) and its derivative during preheating as $M_{\rm eff}=a_eM$ and $M'_{\rm eff}=a_eM'+H_RM$, respectively, where
$a_e=\sqrt{H_R/H_{dS}}$ is the scale factor at the end of inflation.
The mode equation (\ref{Eq_f}) becomes
\begin{equation}\label{Eq_f_tau}
    f_k''+\left( k^2+a_e^2 M^2-i a_e M'-i H_R M \right) f_k = 0 \;.
\end{equation}
This equation admits analytic solution in terms of the hypergeometric function.
The solution is valid in the region $|\eta|\ll \eta_R$.
At $\eta<0$, $|\eta|\gg \tau_\eta$, the solution matches the in-state modes (\ref{f}) corresponding to the Bunch--Davies vacuum.
The matching fixes one parameter in the general solution; another parameter is fixed by the normalisation.
We denote the resulting solution by $\bar{f}_k$.
The out-state modes $f_{\text{out},k}$ with $k\gtrsim k_2$ are in vacuum at a time $\eta_*=\eta_*(k)$ satisfying $\eta_R\gg\eta_*\gg \tau_\eta$.
For such modes one can compute the Bogolyubov coefficient (\ref{BetaMain}) at $\eta=\eta_*$ using $\bar{f}_k$.

The above procedure can be carried out numerically; it is instructive, however, to understand analytically how the quadratic pole of $|\beta_k|^2$ is regularised.
To this end, we focus on cases $(i)$, $(ii)$, and on the limit $\tau_\eta^{-1}\gg k_3$ of case $(iii)$.
In this region of parameters, we find
\begin{equation}\label{BetaTauLargeK}
    |\beta_k|^2_{k_2^*\lesssim k} = \frac{\sh^2{\left( \frac{\pi \tau_\eta k_3}{2} \right)}}{\sh^2{(\pi \tau_\eta k)}} \;,
\end{equation}
where $k_2^*$ is given in \cref{k2*}.
The derivation of this result can be found in Appendix \ref{App:tau>0}.
The expression (\ref{BetaTauLargeK}) replaces the power asymptotics (\ref{BetaInstVarMassHighK}).
It decays exponentially in the limit $k\to\infty$, thus regularising the UV divergence in the particle number and energy densities.
As a consistency check, in the limit of infinitely short preheating \cref{BetaTauLargeK} reduces to \cref{BetaInstVarMassHighK}.
As another check, we notice that the limit $k\gg k_2$ formally corresponds to the limits $H_R,H_{dS}\to 0$, $H_R/H_{dS}=\:$const, or, equivalently, to the limit $M_P\to\infty$ when gravity is switched off.
Then, we can treat \cref{BetaTauLargeK} as describing particle production due to fermion mass variation in the Minkowski background.
This problem was studied in Ref.~\cite{Andreev:2003pg} which uses the same interpolation function (\ref{M_tau}) for the fermion mass.
The result agrees with \cref{BetaTauLargeK} upon rescaling of $M_{i,f}$.

The behavior of $|\beta_k|^2$ at $k\ll k_2$ is identical to the one studied in sec.~\ref{sec:inst}.
To illustrate this, in Fig.~\ref{fig:TauBeta1} we plot the result of the numerical evaluation of $|\beta_k|^2$ and of the particle spectrum $k^2 |\beta_k|^2$, for the values of $M_i$, $M_f$ and $H_{dS}$ corresponding to case $(i)$ and for $\tau_\eta^{-1}\gg k_2$.
These agree with the limit of infinitely short preheating and with the expression (\ref{BetaTauLargeK}) in the respective ranges of comoving momentum.

\begin{figure}[t]
\centering
  \includegraphics[width=0.65\linewidth]{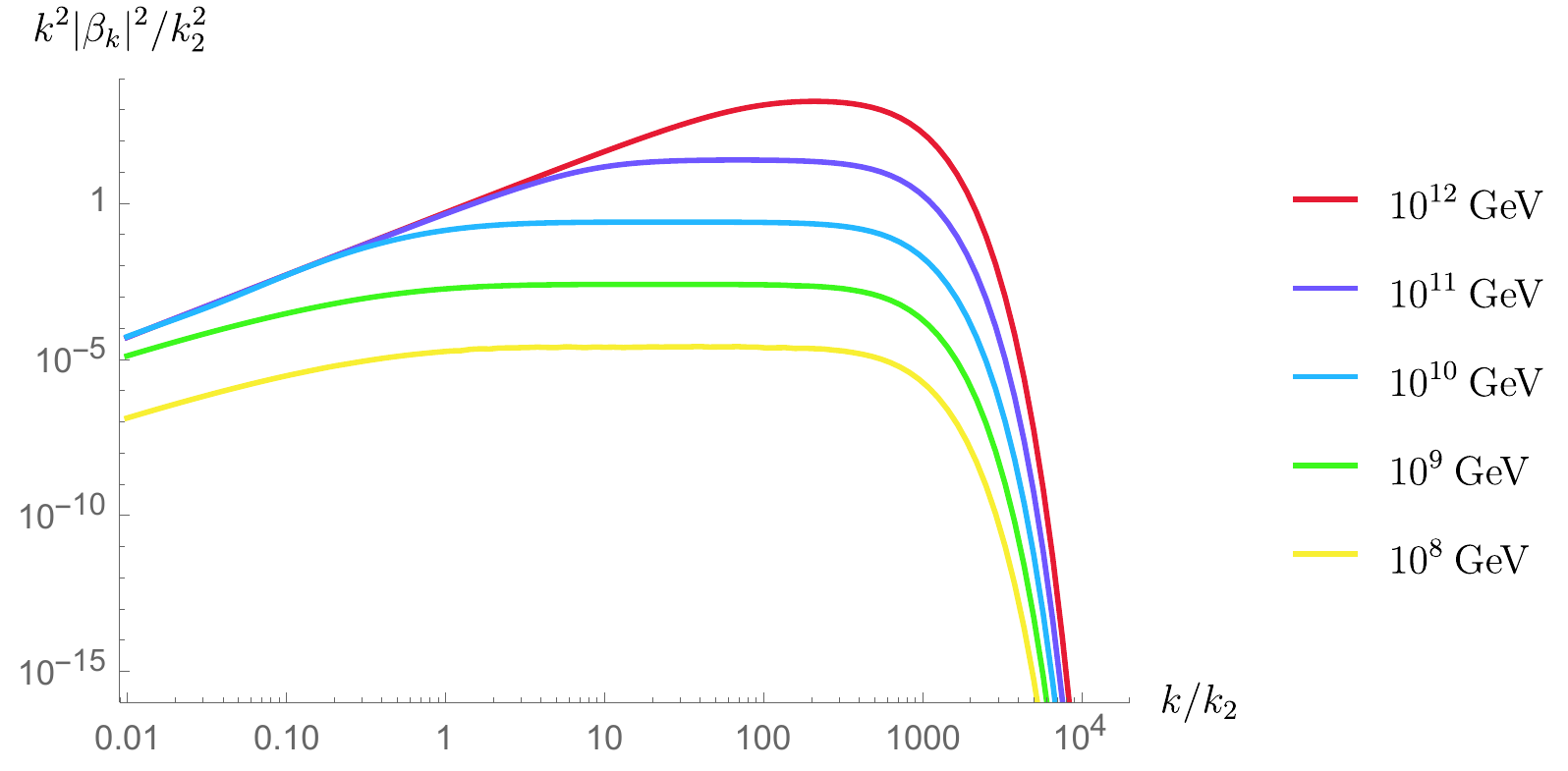}
\caption{Spectral density of fermions produced during preheating, with the time-varying effective fermion mass.
We take $M_f=10^2$ GeV, $H_{dS}=10^{10}$ GeV, $\tau_\eta^{-1}=10^3 k_2$, and several values of $M_i$.
We see that, as long as $\tau_\eta^{-1}\gg k_2^*$, the spectrum is uniform across the modes with $k_2\lesssim k\lesssim \tau_\eta^{-1}$.
On the other hand, when $\tau_\eta^{-1}\lesssim k_2^*$, the modes with $k\sim\tau_\eta^{-1}$ dominate the spectrum.
The scales $k_2$, $k_2^*$ are defined in \cref{k1k2,k2*}, respectively.
}
\label{fig:TauBeta2}
\end{figure}

Let us evaluate the number and energy densities of produced fermions, in the limit $\tau_\eta^{-1}\gg k_2^*$.
The particle spectrum is dominated by energetic modes with the comoving momentum in the range $k_2^*\lesssim k\lesssim \tau_\eta^{-1}$, as one can see from Fig.~\ref{fig:TauBeta2}.
As discussed above, these modes are produced during preheating, and for them \cref{BetaTauLargeK} is applicable.
We substitute it to \cref{n,rho} and evaluate the integrals.
This gives
\begin{equation}\label{n_rho_tau}
    n = \frac{k_3^2 }{12\pi\tau_\eta a^3} \;, ~~~ \rho = M_f n \;,
\end{equation}
where $k_3$ is given in \cref{k3}.
These expressions are accurate provided that $k_2^*/(aM_f)\ll 1$, which is satisfied at the time of matter-radiation equality.\:\footnote{Requiring the DM to be non-relativistic at $a=a_{eq}$ leads to the bound $M_i\ll 10^6 H_{dS}$.}
Substituting to \cref{O1}, we find the particle abundance at $a=a_{eq}$.
Assuming $M_i\gg M_f$ and switching to the physical time, we arrive at
\begin{equation}\label{O3}
     \frac{\Omega_\Psi }{\Omega_{DM}} = \frac{0.3}{\tau  H_{dS}}\left( \frac{H_{dS} }{10^{10}\:\text{GeV}}\right)^{-1/2} \left( \frac{M_i}{10^{10}\:\text{GeV}} \right)^2 \frac{ M_f }{10^5\:\rm{GeV}} \;, ~~ \tau \ll \min ( H_{dS}^{-1}, M_i^{-1} ) \;.
\end{equation}
As expected, the abundance is proportional to the inverse power of the preheating time which we choose to measure in units of $H_{dS}$.

\begin{figure}[t]
\centering
  \includegraphics[width=0.6\linewidth]{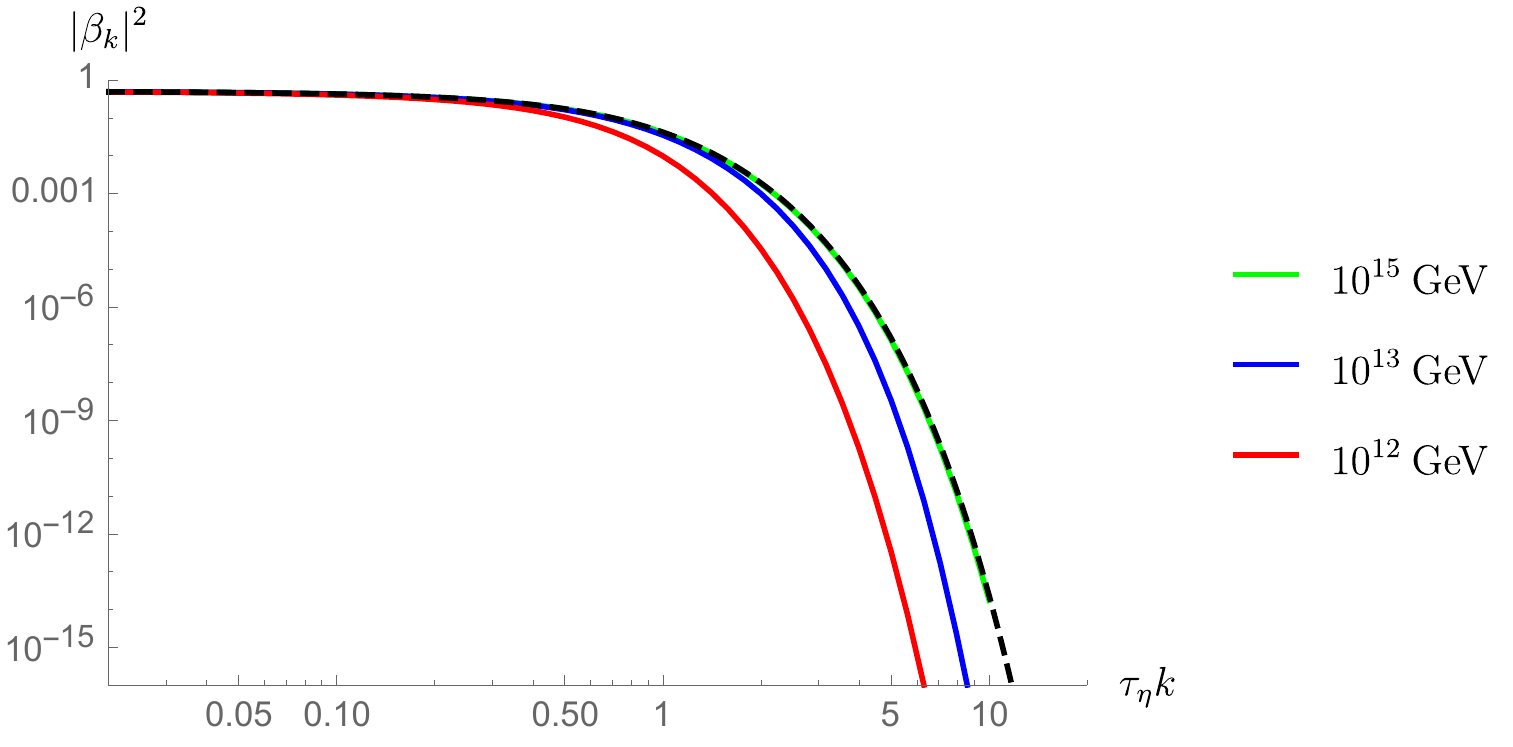}
\caption{Distribution function of fermions produced during preheating in the regime $k_2\ll\tau_\eta^{-1}\lesssim k_3$.
We take $M_f=10^2$ GeV, $H_{dS}=10^{10}$ GeV, $\tau_\eta^{-1}=10^2k_2$ and several values of $M_i$ corresponding to $k_3=\tau_\eta^{-1}$ (red), $10\tau_\eta^{-1}$ (blue), $10^3\tau_\eta^{-1}$ (green).
The dashed line shows the Fermi--Dirac distribution with the temperature (\ref{Teff}).
The scales $k_2$, $k_3$ are defined in \cref{k1k2,k3}, respectively.
}
\label{fig:thermal}
\end{figure}

\begin{figure}[h!]
	\center{
		\begin{minipage}[h]{0.42\linewidth}
			\center{(a) \\ \includegraphics[width=0.99\linewidth]{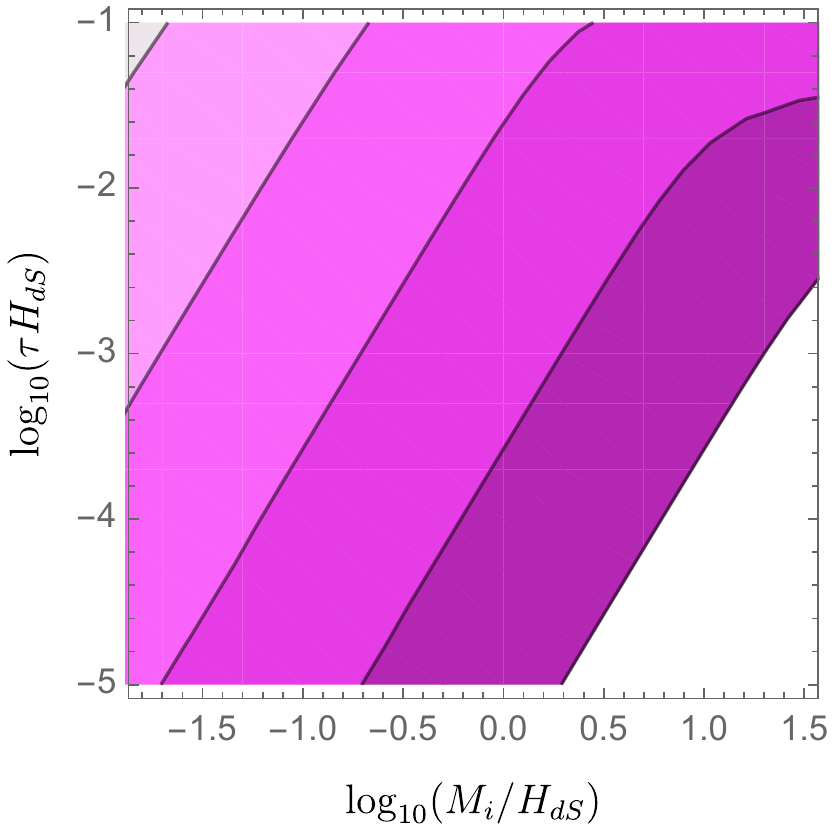}}
		\end{minipage}
		\hfill
		\begin{minipage}[h]{0.42\linewidth}
			\center{(b) \\ \includegraphics[width=0.99\linewidth]{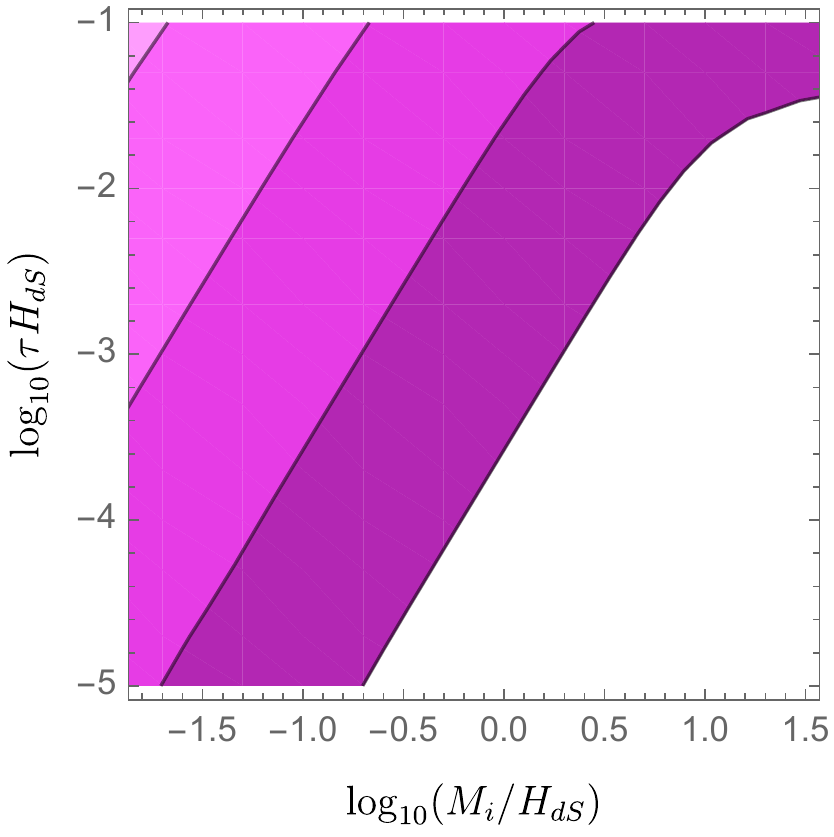}}
		\end{minipage}
		\begin{minipage}[h]{0.05\linewidth}
			\center{$\log_{10}\frac{\Omega_\Psi}{\Omega_{DM}}$ \\ \includegraphics[width=0.99\linewidth]{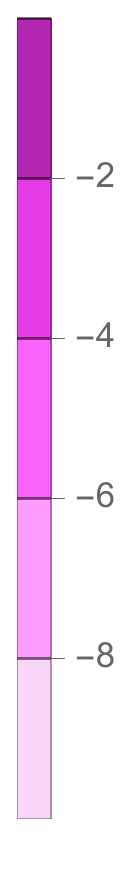}}
		\end{minipage}
		\vfill
		\begin{minipage}[h]{0.42\linewidth}
			\center{(c) \\ \includegraphics[width=0.99\linewidth]{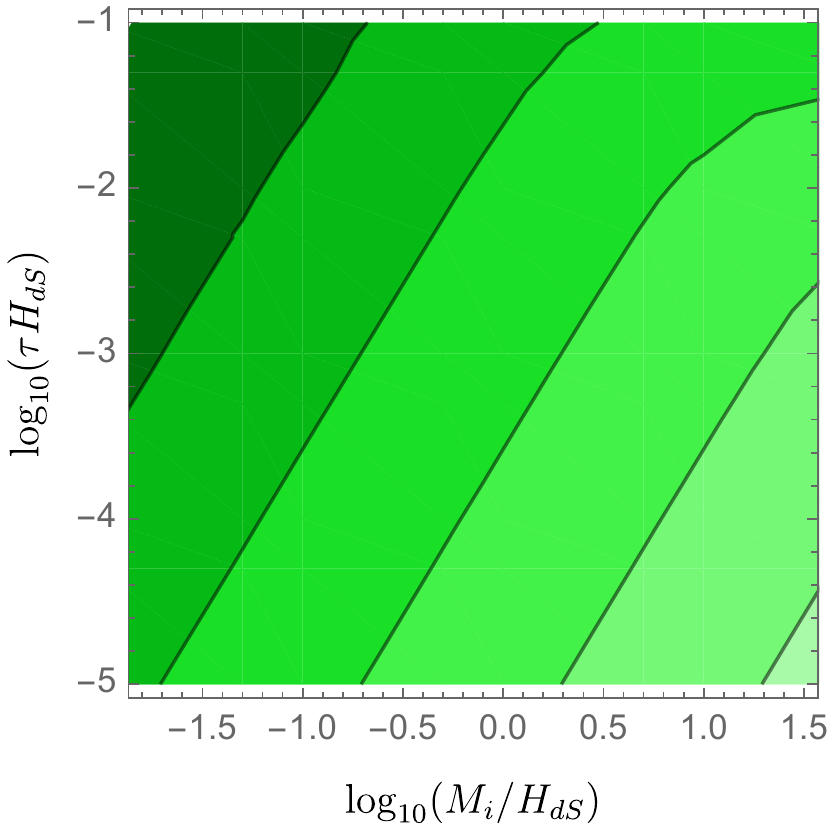}}
		\end{minipage}
		\hfill
		\begin{minipage}[h]{0.42\linewidth}
			\center{(d) \\ \includegraphics[width=0.99\linewidth]{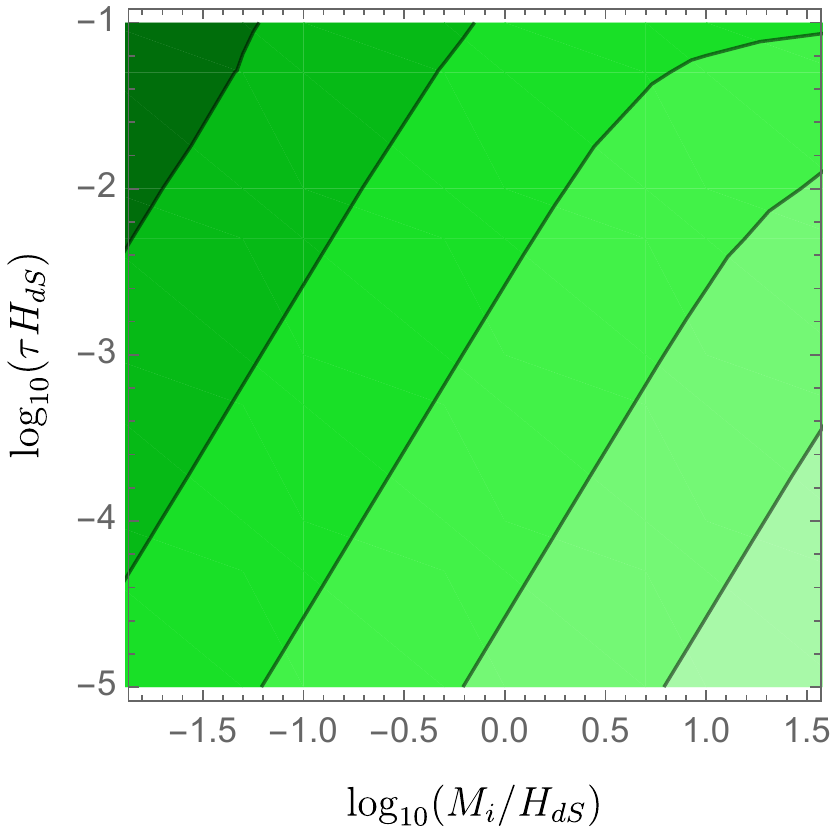}}
		\end{minipage}
		\begin{minipage}[h]{0.05\linewidth}
			\center{$\log_{10}\frac{M_f}{\rm GeV}$ \\ \includegraphics[width=0.99\linewidth]{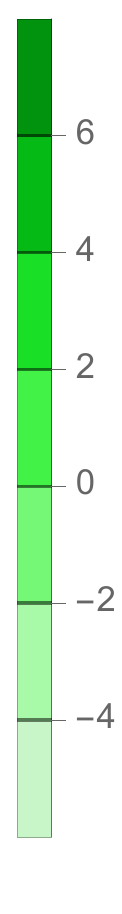}}
		\end{minipage}
	}
	\caption{Abundance $\Omega_\Psi$ of fermions produced during fast (\cref{PhysTau}) and instant (\cref{M_tau}) preheating by the non-perturbative gravitational mechanism assisted by the inflaton-fermion coupling, as a function of the (physical) preheating time $\tau$, the effective fermion mass at the end of inflation $M_i$, and the bare fermion mass $M_f$.
		The plots on top show the contours of constant $\Omega_\Psi$ for $M_f=1$ GeV \textbf{(a)} and $10^2$ GeV \textbf{(b)}.
		The bottom plots show the contours of constant $M_f$ for $\Omega_\Psi/\Omega_{DM}=1$ \textbf{(c)} and $0.1$ \textbf{(d)}.
	}
	\label{fig:TauOmega}
\end{figure}

Let us now explore the case of an initially superheavy fermion, $M_i\gtrsim H_{dS}\gg M_f$ (regime $(iii)$), and the preheating time satisfying $k_2\ll\tau_\eta^{-1}\lesssim k_3$.
In this case, the exponential fall-off of $|\beta_k|^2$ is connected to the $\mathcal{O}(1)$-asymptotics in \cref{BetaInstVarMSmallK} rather than to the power asymptotics (\ref{BetaInstVarMassHighK}).
The particle spectrum is peaked at $k\sim\tau_\eta^{-1}$ (see Fig.~\ref{fig:TauBeta2}), hence the relevant range of comoving momentum is $k_2\ll k\ll k_3$.
In Appendix~\ref{App:tau>0} we show that in this range the particle distribution takes the form
\begin{equation}\label{BetaTauLargeK2}
    |\beta_k|^2_{k_2\ll k\ll k_3} = \frac{1}{1+ \e^{\pi\tau_\eta k} } \;.
\end{equation}
Interestingly, unlike \cref{BetaTauLargeK}, it does not depend on $k_3$, that is on the inflaton-induced fermion mass $M_i$.
It moreover coincides with the Fermi--Dirac distribution with the temperature
\begin{equation}\label{Teff}
    T_{\rm eff}=\frac{1}{\pi\tau } \;,
\end{equation}
where we switched to the physical time.
Thus, in the regime $M_i^{-1}\ll\tau\ll H_{dS}^{-1}$ (and for the particular choice of the interpolation function (\ref{M_tau})), the non-perturbative production mechanism imitates the thermal freeze out of fermions at the temperature $T_{\rm eff}$.
We demonstrate this in Fig.~\ref{fig:thermal}.
Note that $T_{\rm eff}$ is unrelated to the temperature $T_{\rm reh}$ that can possibly be attained at preheating.
Given that the preheating is almost instantaneous, one estimates
\begin{equation}\label{Treh}
    T_{\rm reh}^2=1.66\sqrt{g_{*,\rm{reh}}}M_PH_{dS} \;,
\end{equation}
where $g_{*,\rm{reh}}\sim 10^2$ is the number of relativistic degrees of freedom at that time.
To avoid back-reaction of produced fermions on preheating, one should require $T_{\rm eff}\ll T_{\rm reh}$.
This is, of course, compatible with the condition $\Omega_\Psi/\Omega_{DM}\leqslant 1$.

The particle number and energy densities are
\begin{equation}\label{n_rho_tau2}
    n = \frac{3\zeta(3) }{\pi^5 \tau_\eta^3 a^3} \;, ~~~ \rho = M_f n \;,
\end{equation}
with the same validity condition as for \cref{n_rho_tau}.
From here we obtain the abundance of $\Psi$-particles:
\begin{equation}\label{O4}
    \frac{\Omega_\Psi }{\Omega_{DM}} = \frac{0.1}{( \tau H_{dS})^3} \left( \frac{H_{dS} }{10^{10}\:\text{GeV}}\right)^{3/2}\frac{M_f }{10^5\:\text{GeV}} \;, ~~~ M_i^{-1} \lesssim  \tau \ll H_{dS}^{-1} \;.
\end{equation}
As expected, $\Omega_\Psi$ exhibits enhanced sensitivity to the preheating time, as compared to \cref{O3}.
It is also independent of the inflaton-induced mass $M_i$, that is, of the fermion-inflaton coupling.
At $H_{dS}\tau\sim H_{dS}/M_i$, \cref{O4} qualitatively matches \cref{O3}.

Our findings are summarised in Fig.~\ref{fig:TauOmega}.
It is important to note that the analytic expressions for the particle distribution functions, \cref{BetaTauLargeK,BetaTauLargeK2}, were derived assuming the particular shapes of the scale factor (\ref{a_tau}) and of the inflaton-induced fermion mass (\ref{M_tau}), interpolating between the epochs of inflation and radiation dominated universe.
In a realistic model of preheating, even assuming the condition (\ref{PhysTau}), the solvability of the mode equation (\ref{Eq_f}) is lost.
Nevertheless, we expect that, parametrically, our results for the number density and abundance of particles produced during preheating will remain correct, and in the next section we apply them to a concrete model of inflation and preheating.

\section{Sterile neutrino production in Palatini Higgs Inflation}
\label{sec:HI}

Our goal is to compare the non-perturbative gravitational production mechanism assisted by the fermion-inflaton interaction to the thermal production channel arising from the same interaction.
This requires the inflaton field to be in equilibrium after preheating, as happens in Higgs inflation, in which the role of inflaton is played by the Higgs field $\Phi$~\cite{Bezrukov:2007ep}.
The agreement with cosmological observations is reached by introducing the nonminimal coupling of $\Phi$ to the Ricci curvature scalar.
Since the couplings of the Higgs field to the rest of the Standard Model are known, one can, in principle, calculate the dynamics of preheating, thus connecting inflation to the hot Big-Bang cosmology~\cite{Bezrukov:2008ut,Garcia-Bellido:2008ycs,Bezrukov:2011sz,Enqvist:2013kaa,DeCross:2015uza,Repond:2016sol,Ema:2016dny,DeCross:2016fdz,DeCross:2016cbs,Fu:2017iqg,Sfakianakis:2018lzf,Rubio:2019ypq,He2019,Nguyen:2019kbm,vandeVis:2020qcp,Karam:2020rpa,Hamada:2020kuy,He:2020ivk}.

The original scenario of~\cite{Bezrukov:2007ep} employs the metric formulation of General Relativity, in which the gravitational degrees of freedom are carried by the metric field, and the connection is fixed to be the Levi--Civita one.
Alternatively, Higgs inflation can be studied in the Palatini formulation, in which the metric and (symmetric) connection $\Gamma$ are treated as independent variables \cite{Bauer:2008zj}.\:\footnote{See \cite{Rubio:2018ogq,Tenkanen:2020dge} for reviews of the metric and Palatini Higgs inflation, respectively.}
Depending on matter fields and their coupling to gravity, the metric and Palatini versions of a theory can lead to different predictions.
In Higgs inflation this happens due to the nonminimal coupling of $\Phi$ to gravity.

The dominant mechanism of preheating in Palatini Higgs inflation is the production of Higgs excitations via the tachyonic instability \cite{Rubio:2019ypq,Dux:2022kuk}.
The mechanism is very efficient, and the short preheating time makes the model suitable for our analysis.
Note that the metric Higgs inflation exhibits an explosive production of the longitudinal component of gauge bosons, which takes place at small Higgs values~\cite{DeCross:2016cbs,Ema:2016dny}.
This may invalidate the calculation, since particles produced this way may have energies exceeding the cutoff of the theory~\cite{Bezrukov:2010jz,Gorbunov:2018llf,Bezrukov:2019ylq,Rubio:2019ypq}.
There is no such problem in Palatini Higgs inflation \cite{Rubio:2019ypq}.

Adding the non-minimal coupling of the Higgs field to gravity is an economic extension of the Standard Model and General Relativity that does not require new degrees of freedom to account for inflation.
A further extension of the model amounts to adding heavy right-handed Majorana fermions---sterile neutrino states---two of which can provide masses to light active neutrino and generate baryon asymmetry of the universe, while the third one is a DM candidate~\cite{Asaka:2005pn,Asaka:2005an}.
Being an effective field theory, the model receives corrections in the form of higher-dimensional operators suppressed by a large energy scale \cite{Bezrukov:2008ut,Bezrukov:2011sz}.
Here we consider a 5-dimensional operator that couples the sterile neutrino $\Psi_R$ to the Higgs and gives rise to the large sterile neutrino mass during inflation.

The relevant part of the action of the model takes the form
\begin{equation}\label{ActionJ}
\begin{split}
     S=\int\diff^4 x\sqrt{-g} &\left\lbrace - \frac{1}{2}(M_P^2+\xi \vf^2) g^{\mu\nu}R_{\mu\nu}(\Gamma,\partial\Gamma)+\frac{1 }{2} g^{\mu\nu}\partial_\mu \vf\partial_\nu \vf -\frac{\lambda \vf^4}{4 } \right. \\
    & \left. +i\bar{\Psi}_R\gamma^\mu \mathcal{D}_\mu \Psi_R -\frac{M_R}{2}\bar{\Psi}^C_R\Psi_R-\frac{c_5 }{2\Lambda }\vf^2 \bar{\Psi}^C_R\Psi_R  + \text{h.c.} \right\rbrace
\end{split}
\end{equation}
where we adopted the unitary gauge for the Higgs field, $\Phi=(0,\vf/\sqrt{2})^T$.
Here $\xi$ is the nonminimal coupling, $\l$ is the Higgs quartic self-coupling, $M_R$ is the Majorana mass, $c_5>0$ is the (dimensionless) 5-dimensional coupling.
Next, $\Lambda=\Lambda(\vf)$ is the background-dependent cutoff scale chosen so as to coincide with the cutoff in the gravitational sector of the theory \cite{Bezrukov:2010jz,Bezrukov:2011sz} (see also \cite{Ferrara:2010yw,Shaposhnikov:2020fdv}):
\begin{equation}
    \Lambda(\vf)=\sqrt{M_P^2+\xi\vf^2} \;.
\end{equation}
Finally, in (\ref{ActionJ}) we write the Ricci tensor as a function of $\Gamma$ and its derivative, indicating that in the Palatini formulation of gravity it is a priori independent of the metric.

It is convenient to get rid of the nonminimal coupling by making the Weyl transformation
\begin{equation}
\label{eq:weyl}
g_{\mu\nu}\mapsto\Omega^{-2}g_{\mu\nu} \;, ~~~ \Psi_R\mapsto \Omega^{3/2}\Psi_R \;, ~~~ \Omega^2=1+\frac{\xi \vf^2}{M_P^2} \;.
\end{equation}
Note that the transformation does not involve the connection.
At this point the difference is manifested between the Palatini Higgs inflation and its metric counterpart (see, e.g.,~\cite{Rubio:2019ypq,Shaposhnikov:2020fdv}).
The action (\ref{ActionJ}) becomes
\begin{equation}\label{ActionE}
\begin{split}
S  = \int \diff^{4}x\sqrt{-g} & \left\lbrace -\frac{M_P^2}{2}R+\frac{1}{2\Omega^2}g^{\mu\nu}\partial_{\mu}\vf\partial_{\nu}\vf-\frac{\lambda \vf^{4}}{4\Omega^4}\right. \\
& \left. +i\bar{\Psi}_R\gamma^\mu \mathcal{D}_\mu \Psi_R -\frac{M_R}{2\Omega}\bar{\Psi}^C_R\Psi_R-\frac{c_5 }{2M_P\Omega^2}\vf^2 \bar{\Psi}^C_R\Psi_R  + \text{h.c.} \right\rbrace
\end{split}
\end{equation}
Since the coupling to gravity is now minimal, the connection becomes the Levi--Chivita one, hence we omitted $\Gamma$ in the Ricci scalar $R \equiv g^{\mu\nu}R_{\mu\nu}$.
At large field values, the potential term in \cref{ActionE} flattens and allows for inflation.
The couplings $\xi$ and $\lambda$ at inflationary energy scales are related by the amplitude of primordial spectrum of curvature perturbations~\cite{Planck:2018jri}.
For the benchmark values of these parameters we adopt~\cite{Shaposhnikov:2020fdv}
\begin{equation}\label{values}
\xi=10^7 \;, ~~~ \lambda=10^{-3} \;.
\end{equation}

We now apply the analysis of sec.~\ref{sec:fin} to the theory (\ref{ActionE}).
The Higgs value at the end of inflation is $\vf_{e}\approx M_P$ \cite{Shaposhnikov:2020fdv}.
From the Friedmann equation and the potential term in \cref{ActionE} we obtain the Hubble rate at the end of inflation
\begin{equation}\label{H_ds_HI}
    H_{dS}=\sqrt{\frac{1 }{3M_P^2}\frac{\lambda \vf^4_{e}}{4\Omega(\vf_{e})^4}}=\frac{\sqrt{\lambda } M_P }{2\sqrt{3}\xi } \;,
\end{equation}
where we used the fact that $\xi\gg 1$.
Next, the duration of preheating can be written as
\begin{equation}\label{tau_HI}
    \tau = \frac{2\tilde{t}\xi }{\sqrt{\lambda }M_P } \;,
\end{equation}
where $\tilde{t}$ is a dimensionless parameter.
From \cite{Rubio:2019ypq,Dux:2022kuk} we have $\tilde{t}=0.14$.
On the other hand, from \cref{H_ds_HI,tau_HI} we see that the condition (\ref{PhysTau}) translates into $\tilde t\ll 1$.
Thus, the preheating in Palatini Higgs inflation is fast enough for our analysis to be applicable.

However, one needs to make a small adjustment: we cannot use the ansatz (\ref{M_tau}) for the Higgs-induced fermion mass with the value of $\tau_\eta$ determined by \cref{tau_HI}.
The reason is that, the time $\tau'$ it takes the Higgs field to drop from $\vf\approx\vf_e$ down to $\vf\approx 0$ is approximately 10 times shorter than the total duration of preheating \cite{Rubio:2019ypq,Dux:2022kuk}.
It is easy to see that the scale $\tau$ in the expressions for the DM abundance (\ref{O3}), (\ref{O4}) corresponds to the one in \cref{M_tau}, while the scale in \cref{a_tau} is only needed to establish the condition (\ref{PhysTau}).
Thus, we can simply replace $\tau\mapsto \tau'\approx 0.1\tau$ in \cref{O3,O4} before applying \cref{tau_HI}.

Another comment concerns with the time-evolution of the Higgs profile during preheating.
The analysis of Refs.~\cite{Rubio:2019ypq,Dux:2022kuk} shows that $\vf$ makes one oscillation around the bottom of the potential, before it fragments due to the tachyonic instability rendering the analysis of sec.~\ref{sec:fin} inapplicable.
One can check numerically that the DM production is most efficient during the first half-oscillation of the Higgs field, and its subsequent evolution has a little effect on the resulting DM abundance.
Hence, we can still use the interpolation (\ref{M_tau}) leading to \cref{O3,O4} for the abundance.

From the neutrino mass terms in \cref{ActionE} we read out the initial and final mass parameters,
\begin{equation}\label{Mif_HI}
    M_i=\frac{c_5 M_P}{\xi} \;, ~~~ M_f=M_R \;,
\end{equation}
where we assumed that $M_R\gg c_5 v^2/M_P$ with $v=246$ GeV the today's Higgs vacuum expectation value.
Let us see which of the regimes studied in sec.~\ref{sec:fin} is realised in the model.
An upper bound on the coupling $c_5$ comes from the requirement for the heavy fermion to not spoil inflation~\cite{Bezrukov:2008ut,Bezrukov:2011sz}.
We evaluate the bound in Appendix~\ref{App:Infl} and obtain $c_5\lesssim 10^{-1}$.
Next, from \cref{k1k2,H_ds_HI,tau_HI} we obtain $\tau'H_{dS}\sim 10^{-3}$.
On the other hand, from \cref{Mif_HI} and the inflationary bound on $c_5$ it follows that $\tau'M_i\ll 1$.
Thus, the abundance of produced particles is given by \cref{O3}.
Using \cref{values,H_ds_HI,tau_HI,Mif_HI} we obtain
\begin{equation}\label{O_HI_grav}
    \frac{\Omega_\Psi }{\Omega_{DM}}\sim 3\cdot 10^{9}\frac{c_5^2}{\lambda^{1/4}\xi^{3/2}}\frac{M_R}{1\:\rm GeV} \sim \left(\frac{c_5}{10^{-1}}\right)^2\frac{M_R}{100\:\rm GeV} \;.
\end{equation}

We compare \cref{O_HI_grav} with the abundance resulting from the perturbative production via the same 5-dimensional operator after preheating.
From \cref{Treh,values,H_ds_HI} we find the highest temperature attained at the preheating, $T_{\rm reh}=4\cdot 10^{13}$ GeV.
Then, from the results of Ref.~\cite{Bezrukov:2008ut} it follows that\:\footnote{Ref.~\cite{Bezrukov:2008ut} deals with the metric Higgs inflation, but the difference with the Palatini case disappears at small field values. }
\begin{equation}\label{O_HI_th}
     \frac{\Omega_\Psi }{\Omega_{DM}} = \left(\frac{c_5}{10^{-1}}\right)^2 \frac{M_R}{176\:\rm GeV} \;.
\end{equation}
We see that the non-perturbative channel is as efficient as the perturbative one for the allowed values of $c_5$.

Once fermions are included into consideration, the Palatini Higgs inflation is naturally embedded into the Einstein--Cartan framework~\cite{Langvik:2020nrs,Shaposhnikov:2020gts,Shaposhnikov:2020aen,Shaposhnikov:2020frq,Karananas:2021zkl}.
There, the connection is no longer assumed to be symmetric in its lower indices.
The resulting theory features various nonminimal couplings of scalar and fermion fields to gravity and shows promising implications, e.g., for inflation and DM production~\cite{Langvik:2020nrs,Shaposhnikov:2020gts,Shaposhnikov:2020aen}.
In particular, coupling of fermions to the antisymmetric part of the connection (torsion) leads to effective fermion current-current interaction described by the 6-dimensional operator (see~\cite{Shaposhnikov:2020frq,Karananas:2021zkl} and references therein)
\begin{equation}\label{4f}
   \mathcal{L}'_{int} =  \frac{1}{\Lambda_{VV}}V^\mu V_\mu + \frac{2}{\Lambda_{VA}}V^\mu A_\mu + \frac{1}{\Lambda_{AA}}A^\mu A_\mu \;,
\end{equation}
where $V^\mu = \bar{\Psi}_R \gamma^\mu\Psi_R +\sum_X \bar{X}\gamma^\mu X$ is the vector fermion current, $A^\mu$ is the analogous axial current (with $\gamma^\mu$ replaced by $\gamma^5\gamma^\mu$), and the sum is performed over all Standard Model fermionic species $X$.
The suppression scales $\Lambda_{VV}$, $\Lambda_{VA}$ and $\Lambda_{AA}$ depend on the magnitude of nonminimal couplings of the fermions to gravity.
The thermal production of $\Psi_R$ via the term (\ref{4f}) was studied in \cite{Shaposhnikov:2020aen}, and the resulting abundance can be written as
\begin{equation}\label{O_HI_th2}
     \frac{\Omega_\Psi }{\Omega_{DM}} \sim \left(\frac{M_P}{\Lambda}\right)^4 \frac{M_R}{10^8\:\rm GeV }  \;,
\end{equation}
where $\Lambda$ is the smallest of the suppression scales in (\ref{4f}).
We see that when $\Lambda\sim M_P$, corresponding to the absence of nonminimal fermion couplings, the efficiency of thermal production via the 6-dimensional operator is much lower than of the non-perturbative production via the 5-dimensional operator.
On the other hand, from the analysis of Ref.~\cite{Shaposhnikov:2020gts} it follows that $\Lambda$ can be as small as $M_P/\sqrt{\xi}$ without disrupting inflation.
In this case the four-fermion channel is much more efficient, and a keV-scale sterile neutrino can account for all DM abundance~\cite{Shaposhnikov:2020aen}.

\section{Conclusions}
\label{sec:concl}

We studied non-perturbative production of fermionic dark matter (DM) in the early universe.
Our analysis comprises gravitational production of massive (Dirac or Majorana) fermions combined with the mass-varying effect that can arise during preheating due to the coupling of the fermion to the background inflaton field.
We focused on the case of fast ($\tau H_{dS}\ll 1$) and instant (no oscillations of the inflaton field) preheating and asked how properties of produced fermions are sensitive to it.

In the absence of fermion-inflaton coupling, we reached the known conclusion that fermions are produced abundantly in the evolving curved spacetime during the radiation-dominated era and at $M\sim 10^8$ GeV can account for all observed DM.
The momentum distribution of DM is dominated by modes which were outside the horizon at the time of preheating, thus showing no sensitivity to the details of preheating.

Introducing the fermion-inflaton coupling, in general, leads to drastic variation of the effective fermion mass between the end of inflation and the beginning of radiation-dominated epoch.
The main particle production cite is shifted from the hot Big-Bang to the preheating era.
We modeled the preheating by interpolating analytically the scale factor and the time-varying fermion mass.
We found that the main contribution to the particle spectrum is provided by the modes whose energy is in the range $H_{dS}\lesssim \omega \lesssim \tau^{-1}$, and that their number density scales as $n\propto \tau^{-1}M_i^2$.
Moreover, in the special case of superheavy initial fermion mass, $M_i\gtrsim H_{dS}$ and $M_i \gtrsim \tau^{-1}$, the modes with $\omega\sim\tau^{-1}$ dominate the spectrum.
In this case we also found that the particle distribution coincides with the Fermi--Dirac one of the temperature $T_{\rm eff}=(\pi\tau)^{-1}$ (assuming the interpolation function (\ref{M_tau}) for the time-varying fermion mass).
Accordingly, the number density behaves as $n\propto\tau^{-3}$.
Thus, the ``freeze-in'' of fermionic DM from instant preheating can imitate the freeze-out from thermal plasma with the temperature $T_{\rm eff}$ that is, in general, independent of the preheating temperature.
In all cases, an order-one abundance can be reached in a broad range of parameters.

We illustrated our results using the model of Palatini Higgs inflation, in which the Higgs field plays the role of inflaton and which features the very fast preheating.
We considered a sterile neutrino DM candidate coupled to the Higgs field via the dimension-5 operator.
We found that the non-perturbative gravitational production mechanism assisted by the Higgs-neutrino higher-dimensional coupling can be as efficient as the thermal production via the same operator, for the allowed values of the coupling.

\section*{Acknowledgments}

We thank Mikhail Shaposhnikov for helpful discussions and Inar Timiryasov for participating in the early stage of the project.
The work of A.S. is partially supported by the Department of Energy under Grant No.~DE-SC0011842.
J.K. acknowledges the support of the Fonds de la Recherche Scientifique - FNRS under Grant No.~4.4512.10.

\appendix

\section{Asymptotics of the Bogolyubov coefficient}
\label{App:Beta}

\subsection{\texorpdfstring{$|\beta_k|^2$}{|β|²} in the limit \texorpdfstring{$\tau_\eta=0$}{τ=0}}
\label{App:tau=0}

\subsubsection{Constant fermion mass}
\label{App:constM}

Here we compute the Bogolyubov coefficient, \cref{BetaMain}, in the limits of vanishing fermion-inflaton coupling, $M_i=M_f=M$, and infinitely short preheating, $\tau_\eta=0$.
The latter allows us to identify $\bar{f}_k$ with $f_{\text{in},k}$ at $\eta=0$.
Denote
\begin{equation}\label{Eps}
    \epsilon=\sqrt{\frac{M}{H_{dS}}}=\frac{k_1}{k_2} \;.
\end{equation}

\paragraph{The case $\epsilon\ll 1$.}
We start with the fermion mass small compared to the Hubble rate at the end of inflation.
Consider first the range of momenta $k\ll k_2$.
Then one can derive an analytical expression for $|\beta_k|^2$.
From \cref{etaR} we see that $k\ll k_2$ is equivalent to $k\eta_R\ll 1$, and the inflationary modes (\ref{f}) at $\eta=0$ can be expanded as
\begin{equation}\label{fInExp}
    f_{\text{in},k}(0)=\e^{ik\eta_R}\sum_{n=0}^\infty\frac{a_n }{(k\eta_R)^n} \;, ~~~ a_0 = 1 \;.
\end{equation}
It suffices to take the plane-wave approximation for the modes (\ref{f}).
On the other hand, the argument of the out-state modes $f_{\text{out},k}$ from \cref{f_RD} at $\eta=0$ is proportional to $\epsilon$, and we can expand them to the leading order in $\epsilon$.
Substituting these expansions to \cref{BetaMain} and using \cref{Gamma_eq}, we obtain
\begin{equation}\label{Beta_A1}
    |\beta_k|^2 = \frac{1}{2} + \frac{iq\e^{-\frac{\pi q^2 }{4}}}{4} \left[ \frac{ \e^{\frac{i\pi }{4}}}{\Gamma\left(\frac{1}{2} + \frac{iq^2}{4}\right)\Gamma\left(1- \frac{iq^2}{4}\right)} - \frac{\e^{-\frac{i\pi }{4}}}{\Gamma\left(\frac{1}{2} - \frac{iq^2}{4}\right)\Gamma\left(1+ \frac{iq^2}{4}\right)}\right] \; ,
\end{equation}
where we denoted $q = k/k_1$.
Next, we apply \cref{Gamma_eq2} to $\Gamma\left(\frac{1}{2} \pm \frac{iq^2}{4}\right) $ in the expression (\ref{Beta_A1}), and rewrite the resulting products of Gamma functions using the fact that $z=|z|\e^{i\Im\ln z}$ for any complex $z$.
This way we arrive at\:\footnote{Ref.~\cite{Herring:2020cah} misses $\cos\Phi(k)$ in \cref{Beta_A1_2}.}
\begin{equation}\label{Beta_A1_2}
   |\beta_k|^2 =\frac{1}{2}\left(1-\sqrt{1 - \e^{-\pi q^2}}  \cos{\Phi(q)}\right) \; ,
\end{equation}
where
\begin{equation}
    \Phi(q) = \Im{\left[ \ln \Gamma\left(\frac{-iq^2}{2}\right)- \ln\Gamma\left(\frac{-iq^2}{4 }\right) +\ln\Gamma\left(\frac{iq^2}{4}\right)   \right]} + \frac{\pi}{4} + \frac{1}{2} \ln 2q^2 \; .
\end{equation}
Expanding \cref{Beta_A1_2} in the limits $q\ll 1$ and $q\gg 1$, one recovers the first and the second of the asymptotics (\ref{BetaInstSmallM}), respectively.

Consider now the modes with $k\gg k_2$, which are inside the horizon at the end of inflation.
To handle this case, we need the expansion of the parabolic cylinder function in \cref{f_RD} at large absolute values of its order.
Using \cref{f_RD,U_exp,U1_exp,g_exp,xi(t),Ag,UtoD}, we obtain up to $\mathcal{O}(q^{-2})$
\begin{equation}\label{f_rd_2}
\frac{f_{\text{out},k}'(0)}{f_{\text{out},k}(0)} = - \e^{\frac{i\pi }{4}}\:k_1\: \sqrt{1+ i  \epsilon^2 +i q^2} \: \left( 1-\frac{i  \epsilon}{2q^3} \right) \;.
\end{equation}
Next, we apply the expansion of the inflationary mode function (\ref{fInExp}) and its derivative at $\eta=0$ up to $\mathcal{O}(\eta_R^{-3}k^{-3})$ and $\mathcal{O}(\eta_R^{-2}k^{-2})$, respectively.
Substituting these expansions and \cref{f_rd_2} to \cref{BetaMain}, we arrive at
\begin{equation}\label{Beta_A1_3}
  |\beta_k|^2 = \frac{\epsilon^4}{16\eta_R^6 k^6} \;.
\end{equation}
Expanding to the leading order in $\epsilon$ and using \cref{etaR,k1k2,Eps}, we get the third asymptotics in \cref{BetaInstSmallM}.

\paragraph{The case $\epsilon\gg 1$.}
Consider now the case of a superheavy fermion.
Note that \cref{Beta_A1_3} is exact in $\epsilon$, hence in the high-momentum limit, $k\gg k_1$, we have $|\beta_k|^2\propto k^{-6} $.
When $k\ll k_2$, we use low-momentum asymptotics for the both mode functions and find
\begin{equation}
    |\beta_k|^2_{k\lesssim k_0} = \e^{-2 \pi \epsilon^2} \;, ~~~ |\beta_k|^2_{k_0\lesssim k \lesssim k_2}=\frac{k^2}{16 \epsilon^{10} k_1^2} \;, ~~~ k_0=4k_1\epsilon^5\:\e^{-\pi\epsilon^2} \;.
\end{equation}
Finally, we see numerically that the $k^2$-asymptotics persists up until $k\sim k_1$.
These asymptotics agree with the previous studies; see, e.g., Fig.~1 in \cite{Chung:2011ck}.
We conclude that production of superheavy fermions is suppressed at the moment of preheating.

\subsubsection{Varying fermion mass}
\label{App:varM}

We turn to the case when the inflaton-fermion coupling provides $\Psi$ with the large mass $M_i$ during inflation.
As discussed in the main text, we work in the regime $M_i\geqslant M_f$ and $H_{dS}\gg M_f$.
This implies $k_2=\eta_R^{-1}\gg k_1$, where $k_1$ is defined in \cref{k1new}.
Denote
\begin{equation}\label{EpsIF}
    \epsilon_{i,f}=\sqrt{\frac{M_{i,f}}{H_{dS}}} \;.
\end{equation}

\paragraph{High-momentum asymptotics.}
The behavior of $|\beta_k|^2$ in the large-$k$ limit is determined by the term proportional to the mass difference in \cref{BetaMain}.
For the out-state modes we take the leading in $k$ term in the expansion (\ref{f_rd_2}).
For the in-state modes we adopt the plane-wave approximation.
Both approximations require that $k\eta_R\gg 1$.
Furthermore, from the series (\ref{fInExp}) it follows that the sufficient condition of applicability of the plane-wave approximation is $k\eta_R\gg \epsilon_i^4$.
This is because the coefficients $a_n$ are polynomials in $\epsilon_i$ of degree $4n$.
However, this condition can be relaxed by using the properties of the asymptotic expansion of the Hankel function at large absolute values of its order.
We summarise these properties in \cref{H_exp,W_exp,P2,zetaZ}.
Using also \cref{H_cont} we obtain that it is sufficient to require $k\eta_R\gg \epsilon_i^2$.
Thus, we arrive at \cref{BetaInstVarMassHighK,k2*}.

\paragraph{Low-momentum asymptotics.}

It is clear that the behavior of $|\beta_k|^2$ at low momenta, $k\eta_R\ll 1$, is identical to that in the constant-mass case, provided that $\epsilon_i\ll 1$ (regimes $(i)$, $(ii)$; see Fig.~\ref{fig:InstBetaM1}).
On the other hand, the case of superheavy initial fermion mass, $\epsilon_i\gg 1$, requires a separate study.
Assume that
\begin{equation}\label{EpsEps}
    \epsilon_f\gg\epsilon_i^2\e^{-\pi\epsilon_i^2} \;.
\end{equation}
For $\epsilon_i\geqslant \mathcal{O} (10)$ this assumption translates to $M_f\gg 10^{-26}H_{dS}$, which is always satisfied for the DM candidate \cite{Tremaine:1979we,Planck:2018jri}.
We expand $f_{\text{in},k}(0)$ and $f_{\text{in},k}'(0)$ to first order in $k\eta_R$.
Next, we expand $f_{\text{out},k}(0)$ and $f_{\text{out},k}'(0)$ to leading order in $q$ and first order in $\epsilon_f$ for $k\ll k_1$, or take the leading in $q$ term in the expansion (\ref{f_rd_2}) (with $\epsilon$ replaced by $\epsilon_f$) for $k_1\ll k\ll k_2$.
This way we obtain
\begin{equation}\label{BetaVarMassSmallK2}
    |\beta_k|^2_{k\lesssim k_0}=\epsilon_i^4\e^{-2\pi\epsilon_i^2} \;, ~~~  |\beta_k|^2_{k_0\lesssim k\lesssim k_1}=\frac{\pi k^2 }{4k_1^2} \;, ~~~ |\beta_k|^2_{k_1\lesssim k\lesssim k_2}=\frac{1 }{2}  \;, ~~~ k_0=k_2\epsilon_i^2\e^{-\pi\epsilon_i^2} \;,
\end{equation}
where $k_0\ll k_1$ in view of \cref{EpsEps}.
Note that there is no smooth interpolation between the first and the second asymptotics in \cref{BetaVarMassSmallK2}, as one can also see from Fig.~\ref{fig:InstBetaM2}.
Note also that the exact behavior of $|\beta_k|^2$ at $k\lesssim k_0$ is of no physical consequence, and we can safely replace it by the quadratic asymptotics, which yields \cref{BetaInstVarMSmallK}.

\subsection{\texorpdfstring{$|\beta_k|^2$}{|β|²} in the regime \texorpdfstring{$0<\tau_\eta\ll \eta_R$}{0<τ\_η<<η\_R}}
\label{App:tau>0}

\subsubsection{The case \texorpdfstring{$k_2^*\ll\tau_\eta^{-1}$}{k₂<<τ\_η \^\ (-1)}}

Here we compute the high-momentum asymptotics (\ref{BetaTauLargeK}) of the Bogolyubov coefficient with the finite duration of preheating and assuming that $\tau_\eta^{-1}\gg k_2^*$.
In the region $|\eta|\ll\eta_R$, the mode equation (\ref{Eq_f}) takes the form (\ref{Eq_f_tau}).
Its general solution is
\begin{equation}\label{GenSol}
\begin{split}
    & f_k=c_1 \e^{-i \Tilde{E_i} \eta}(1+ \e^{\frac{2 \eta}{\tau_\eta}})^{i\Delta}\:_2F_1\left( -\frac{i \tau_\eta(\Tilde{E_f} + \Tilde{E_i})}{2} + i\Delta, \frac{i \tau_\eta(\Tilde{E_f}  -\Tilde{E_i})}{2} + i\Delta, 1- i \tau_\eta \Tilde{E_i}; -\e^{\frac{2 \eta}{\tau_\eta}}\right) \\
    & + c_2 \e^{i \Tilde{E_i} \eta}(1+ \e^{\frac{2 \eta}{\tau_\eta}})^{i\Delta}\:_2F_1\left(-\frac{i \tau_\eta(\Tilde{E_f} - \Tilde{E_i})}{2} + i\Delta, \frac{i \tau_\eta(\Tilde{E_f}  +\Tilde{E_i})}{2} + i\Delta, 1+ i \tau_\eta \Tilde{E_i}; -\e^{\frac{2 \eta}{\tau_\eta}}\right) \;.
\end{split}
\end{equation}
Here $_2F_1$ is the hypergeometric function, $c_{1,2}$ are constants, and we denoted
\begin{equation}\label{Consts}
 \Delta = \frac{\tau_\eta (\tilde{M}_f - \tilde{M}_i)}{2} \qquad   \Tilde{E}_{i,f}= \sqrt{k^2 + \tilde{M}^2_{i,f}-i H_R M_{i,f} }, \qquad \tilde{M}_{i,f} = \sqrt{\frac{H_R}{H_{dS}}} M_{i,f} \;.
\end{equation}
To find $c_{1,2}$, one should match this solution to the in-state modes $f_{\text{in},k}$.
The matching is performed in the region $\eta<0$, $\eta_R\gg|\eta|\gg\tau_\eta$.
In this region, the in-state modes are approximated by the plane waves, $f_{\text{in},k}=\e^{-ik(\eta-\eta_R)}$.
On the other hand, the solution (\ref{GenSol}) becomes the linear combination of plane waves, $f_k=c_1\e^{-i k\eta}+c_1\e^{ik\eta}$, where we assumed that $k\gg k_2$ for $M_i\ll H_{dS}$ (cases $(i)$, $(ii)$) or that $k\gg k_2 M_i/H_{dS}$ for $M_i\gtrsim H_{dS}$ (case $(iii)$).
Thus, the matching requires $c_1=\e^{ik\eta_R}$, $c_2=0$.

Consider now the region $\eta>0$, $\eta_R\gg|\eta|\gg\tau_\eta$, where the Bogolyubov coefficient is evaluated.
The out-state modes satisfy $f_{\text{out},k}\propto\e^{-ik\eta}$, as one can see from the large-$k$ expansion of \cref{f_RD} using \cref{U_exp,g_exp,xi(t),Ag,UtoD}.
To find the behaviour of the solution $\bar{f}_k$, we rewrite the hypergeometric function in (\ref{GenSol}) using \cref{Hyper}, and expand at small values of the argument.
We find $\bar{f}_k=c_3\e^{ik\eta}+c_4\e^{-ik\eta}$ where
\begin{equation}
 c_3=\e^{ik\eta_R}\frac{\Gamma\left(  1- i \tau_\eta k \right)\Gamma\left( i\tau_\eta k \right)}{\Gamma\left( i\Delta \right)\Gamma\left(1 - i\Delta \right)} \;, ~~~ c_4=\e^{ik\eta_R} \;,
\end{equation}
and we used the same conditions on $k$ as in the previous region.
The Bogolyubov coefficient is given by
\begin{equation}
    |\beta_k|^2_{k^*_2\lesssim k}=|c_3|^2=\frac{\sh^2(\pi\Delta)}{\sh^2(\pi\tau_\eta k)} \;,
\end{equation}
where $k^*_2$ is given in \cref{k2*}.
Redefining the variables according to \cref{k3,Consts}, we obtain (\ref{BetaTauLargeK}).

\subsubsection{The case \texorpdfstring{$k_2\ll\tau_\eta^{-1}\ll k_3$}{k₂ << τ\_η \^\ (-1) << k₃}}

Now we consider the case when $\tau_\eta^{-1}$ is not the largest scale in game,
and derive the Bogolyubov coefficient in the range of momenta satisfying $k_2\ll k\sim\tau_\eta^{-1}\ll k_3$.
First, we need the expansion of the in-state modes (\ref{f}) in the regime $\epsilon_i\gg 1$, $k\eta_R/\epsilon_i^2\equiv\delta_i^2\ll 1$, where $\epsilon_i$ is given in \cref{EpsIF}.
We use \cref{HtoJ,JtoF} to express the Hankel function in \cref{f} via the hypergeometric function $_0 F_1$, and expand the latter at small values of $\delta_i$.
This way we obtain $f_{\text{in},k}=c_1\e^{-i k_3\eta}+c_2\e^{ik_3\eta}$ where
\begin{equation}\label{c12}
    c_1=(k\eta_R)^{1+i\epsilon_i^2}\epsilon_i^{-2-i\epsilon_i^2}\e^{i\epsilon_i^2}2^{-\frac{1}{2}-i\epsilon_i^2} \;, ~~~ c_2=0\;.
\end{equation}
This matches the solution (\ref{GenSol}) of the mode equation (\ref{Eq_f}) in the region $\eta<0$, $\eta_R\gg |\eta|\gg\tau_\eta$.
Continuing to the out-state matching region, $\eta_R\gg \eta \gg \tau_\eta$, we find $\bar{f}_k=c_1c_3\e^{ik\eta}+c_1c_4\e^{-ik\eta}$ where
\begin{equation}\label{c34}
    c_3=\frac{\Gamma(i\tau_\eta k)\Gamma(1-i\tau_\eta k_3)}{\Gamma\left( \frac{i}{2}\tau_\eta (k-2k_3)\right)\Gamma\left(1+\frac{i\tau_\eta k}{2} \right)} \;, ~~~ c_4=\frac{\Gamma(-i\tau_\eta k)\Gamma(1-i\tau_\eta k_3)}{\Gamma\left( -\frac{i}{2}\tau_\eta (k+2k_3)\right)\Gamma\left(1-\frac{i\tau_\eta k}{2} \right)} \;,
\end{equation}
and we retained the subleading in $\tau_\eta k_3$ terms.
The Bogloyubov coefficient is given by
\begin{equation}
    |\beta_k|^2_{k_2\ll k\ll k_3}=|c_1|^2|c_3|^2  \;,
\end{equation}
and, using \cref{c12,c34}, we arrive at \cref{BetaTauLargeK2}.

\section{Contribution of the higher-dimensional term to inflation}
\label{App:Infl}

At tree level, the Higgs potential during inflation is written as
\begin{equation}\label{U0}
    V(\vf)=\frac{\l \vf^4}{4\Omega^4}\approx \frac{\l M_P^4 }{4\xi^2}-\frac{\l M_P^6 }{2\xi^3 \vf^2} \;,
\end{equation}
where we used \cref{eq:weyl} and the fact that $\xi \vf^2/ M_P^2\gg 1$ during inflation in the Palatini model.
The one-loop contribution of the 5-dimensional operator in \cref{ActionE} to this potential is
\begin{equation}\label{dU}
      \delta V(\vf)=-\frac{M(\vf)^4 }{32\pi^2}\log\frac{M(\vf)^2}{\mu_{\rm inf}^2} \;,
\end{equation}
where $M(\vf)$ is the mass of $\Psi_R$ provided by the 5-dimensional operator,
\begin{equation}\label{M(h)}
    M(\vf)=\frac{c_5 \vf^2}{M_P\Omega^2}\approx \frac{c_5M_P}{\xi} \;,
\end{equation}
and we neglected the bare fermion mass $M_R$.
The inflationary energy scale $\mu_{\rm inf}$ is of the order of the top quark mass during inflation
\begin{equation}\label{MuInf}
    \mu_{\rm inf}=y_t\frac{M_P}{\sqrt{\xi}} \;,
\end{equation}
where $y_t=0.43$ is the value of the top Yukawa coupling during inflation, which corresponds to the value $\lambda=10^{-3}$ of the Higgs quartic coupling during inflation, see \cite{Shaposhnikov:2020fdv} for details.

The contribution (\ref{dU}) does not spoil inflationary predictions if the change in the slope of the inflationary potential caused by the correction is small at $N_* \approx 50$ e-folds,\:\footnote{These conditions follow from $\frac{\diff}{\diff \chi}(V+\delta V)>0$ and from the fact that $\frac{\diff V}{\diff \chi}\propto \frac{\diff V }{\diff \vf}$, where $\chi$ is the canonically normalised scalar field, which is a monotonic function of $\vf$ \cite{Shaposhnikov:2020fdv}.  }
\begin{equation}
    \frac{\diff }{\diff \vf}(V+\delta V)>0 \;, ~~~  \left\vert \frac{\diff \delta V }{\diff \vf}\right\vert \lesssim \frac{\diff V }{\diff \vf} \;, ~~~ M_P\lesssim \vf\lesssim M_P\sqrt{N_*} \;.
\end{equation}
Using \cref{U0,dU,M(h),MuInf,values} we obtain $c_5 \lesssim   10^{-1}$.

\section{Some useful formulas}
\label{App:Eqs}

Properties of the Gamma function:
\begin{equation}\label{Gamma_eq}
    \left\vert\Gamma(1+ix)\right\vert^2 = \frac{\pi x}{\sh{\pi x}} \;, ~~~ \left\vert\Gamma\left(\frac{1}{2}+ix\right)\right\vert^2 = \frac{\pi }{\ch{\pi x}} \;,
\end{equation}
where $x$ is real;
\begin{equation}\label{Gamma_eq2}
    \Gamma(2z)=\frac{1}{\sqrt{2\pi }}2^{2z+\frac{1 }{2} }\Gamma(z)\Gamma\left( z +\frac{1 }{2 }\right) \;,
\end{equation}
for any $z$.

Asymptotic expansions of the parabolic cylinder function and its derivative at large $|\mu|$ and $|\arg(\mu)| < \pi$ \cite{Olver1959UniformAE}:
\begin{align}
& U\left(-\frac{1}{2}\mu^2, \mu t \sqrt{2}\right)  \sim g(\mu) \frac{\e^{- \mu^2 \xi }}{(t^2-1)^{\frac{1}{4}}} \sum_{s=0}^{\infty} \frac{A_s(\xi)}{\mu ^{2 s}} \;, \label{U_exp} \\
& U'\left(-\frac{1}{2}\mu^2, \mu t \sqrt{2}\right) \sim - \frac{\mu}{\sqrt{2}} g(\mu) (t^2-1)^{\frac{1}{4}} \e^{- \mu^2 \xi } \sum_{s=0}^{\infty} \frac{B_s(\xi)}{\mu ^{2 s}} \;,  \label{U1_exp}
\end{align}
where
\begin{equation}\label{g_exp}
    \frac{1}{g(\mu)}  \sim 2^{\frac{1}{4}\mu^2 + \frac{1}{4}} \e^{\frac{1}{4}\mu^2} \mu^{-\frac{1}{2}\mu^2 +\frac{1}{2}} \sum_{s=0}^{\infty} \frac{g_s}{\mu^{2 s}} \;,
\end{equation}
and
\begin{equation}\label{xi(t)}
    \xi = \frac{1}{2} t \sqrt{t^2-1} -\frac{1}{2}\ln \left[ t + \sqrt{t^2-1} \right] \;.
\end{equation}
The first coefficients in the series in \cref{U_exp,U1_exp,g_exp} are
\begin{equation}\label{Ag}
    A_0 = B_0 = g_0 = 1 \;, ~~~ A_1 = \frac{t^3 -6t}{24(t^2-1)^{\frac{3}{2}}} \;, ~~~ B_1 = \frac{t^3 +6t}{24(t^2-1)^{\frac{3}{2}}} \;, ~~~ g_1 = \frac{1 }{24} \;.
\end{equation}
The function $U(a,x)$ is related to $D_\alpha(x)$ as follows
\begin{equation}\label{UtoD}
    U(a,x)=D_{-a-\frac{1 }{2}}(x) \;.
\end{equation}

Asymptotic expansion of the Hankel function of the first kind at large $|\nu|$ and $|\arg(\nu)|<\frac{\pi}{2}$, $\arg(z)>0$ \cite{Olver1954a,Olver1954b}:
\begin{equation}\label{H_exp}
    H_\nu^{(1)}(\nu z) \sim \frac{2\e^{-\frac{i\pi }{3}}}{\nu^{\frac{1}{3}}}\left( \frac{4\zeta }{1-z^2} \right)^{\frac{1}{4}}W_2(\zeta) \;,
\end{equation}
where
\begin{equation}\label{W_exp}
    W_2(\zeta)\sim P_2\left( \nu^{\frac{2}{3}}\zeta \right)\sum_{s=0}^\infty \frac{C_s(\zeta)}{\nu^{2s}} + \frac{P_2'\left( \nu^{\frac{2}{3}}\zeta \right)}{\nu^{\frac{4}{3}}}\sum_{s=0}^\infty \frac{D_s(\zeta)}{\nu^{2s}}
\end{equation}
and
\begin{equation}\label{P2}
    P_2(u)=\text{Ai}\left(u\e^{\frac{2\pi i}{3}}\right) \;.
\end{equation}
If $|\zeta|\to\infty$ and $|\arg(-\zeta)|<\frac{2\pi }{3}$, then $|z|\to\infty$ and
\begin{equation}\label{zetaZ}
    \zeta\sim -\left(\frac{3}{2}z\right)^{\frac{2}{3}} \;.
\end{equation}
In this case, the first coefficients in the series in \cref{W_exp} are $C_0=1$, $D_0=\mathcal{O}\left( |z|^{-\frac{4}{3}} \right)$.

Continuation formula for the Hankel function:
\begin{equation}\label{H_cont}
    H_\nu^{(1)}\left(\nu z \e^{i\pi n}\right) = \e^{i\pi n \nu} H_\nu^{(1)}(\nu z) \;,
\end{equation}
$n$ being any integer.

Relation between the Hankel and the Bessel functions (eq.~(10.4.7) in \cite{NIST}):
\begin{equation}\label{HtoJ}
    H_\nu^{(1)}(z)=\frac{i}{\sin\pi\nu}(\e^{-i\pi\nu}J_\nu(z)-J_{-\nu}(z)) \;.
\end{equation}

Relation between the Bessel function and the hypergeometric function $_0F_1$ (eq.~(6.2.7(1)) in \cite{1969special}):
\begin{equation}\label{JtoF}
    J_\nu(z)=\left(\frac{z}{2}\right)^\nu\frac{1}{\Gamma(\nu+1)}\:_0F_1\left(\nu+1,-\frac{z^2}{4}\right) \;.
\end{equation}

Transformation of variables in the hypergeometric function (eqs.~(15.8.2), (15.1.2) in \cite{NIST}):
\begin{equation}\label{Hyper}
\begin{split}
\frac{\sin(\pi(b\!-\!a))}{\pi \Gamma(c)} ~_2F_1 (a,b,c;z)
&=\frac{(-z)^{-a}}{\Gamma(b)\Gamma(c\!-\!a)\Gamma(a\!-\!b\!+\!1)}
~_2F_1 \bigg(a,a\!-\!c\!+\!1,a\!-\!b\!+\!1;\frac{1}{z}\bigg) \\
&-\frac{(-z)^{-b}}{\Gamma(a)\Gamma(c\!-\!b)\Gamma(b\!-\!a\!+\!1)}
~_2F_1 \bigg(b,b\!-\!c\!+\!1,b\!-\!a\!+\!1;\frac{1}{z}\bigg) \;,
\end{split}
\end{equation}
assuming $|\arg(-z)|<\pi$.

{\small
\bibliographystyle{utphys}
\bibliography{refs}
}

\end{document}